\documentclass[12pt,nohyper,notoc]{JHEP}

\usepackage{amsmath,amssymb,cite} 
	
\DeclareSymbolFont{AMSb}{U}{msb}{m}{n}
\DeclareSymbolFontAlphabet{\Bbb}{AMSb}

\def\rsen{\setcounter{equation}{0}}
\newcommand{\startappendix}{
\setcounter{section}{0}
\renewcommand{\thesection}{\Alph{section}}
\renewcommand{\theequation}{\Alph{section}.\arabic{equation}}}
\newcommand{\Appendix}[1]{
\refstepcounter{section}
\begin{flushleft}
{\large\bf Appendix \thesection: #1}
\end{flushleft}}

\newcommand{\aD}{{\dot\alpha}}

\newcommand{\bD}{{\dot\beta}}
\newcommand{\U}{\text{U}}
\newcommand{\SU}{\text{SU}}
\newcommand{\SO}{\text{SO}}

\newcommand{\BL}{{\bf L}}

\newcommand{\BN}{{\boldsymbol{N}}}
\newcommand{\BR}{{\boldsymbol{R}}}
\def\com{X}

\def\sfc{\hat\Omega}

\def\N{{\cal N}}
\def\det{{\rm det}}

\def\M{{\cal M}}
\def\tr{{\rm tr}}
\def\trtwo{\tr^{}_2\,}
\def\Mbar{\bar{\cal M}}
\def\dalpha{{\dot\alpha}}
\def\dbeta{{\dot\beta}}

\def\sqrtwo{\sqrt{2}\,}
\def\hf{{\textstyle{1\over2}}}
\def\wbar{\bar w}
\def\mubar{\bar\mu}
\def\abar{\bar a}

\def\etabar{\bar\eta}

\def\mubar{\bar\mu}

\def\N{{\cal N}}

\def\sst{\scriptscriptstyle}

\def\Mbar{\bar{\M}}

\def\D{{\cal D}}
\def\Dbarslash{\,\,{\raise.15ex\hbox{/}\mkern-12mu {\bar\D}}}
\def\delslash{\,\,{\raise.15ex\hbox{/}\mkern-9mu \partial}}
\def\Dslash{\,\,{\raise.15ex\hbox{/}\mkern-12mu \D}}

\def\mubar{\bar\mu}
\def\dalpha{{\dot\alpha}}

\def\etabar{\bar\eta}

\def\N{{\cal N}}
\def\M{{\cal M}}
\def\A{{\cal A}}
\def\hf{{\textstyle{1\over2}}}

\def\dbeta{{\dot\beta}}
\def\abar{\bar a}
\def\wbar{\bar w}
\def\trtwo{\tr^{}_2\,}

\def\bigL{{\bf L}}

\def\sqrtwo{\sqrt{2}\,}

\title{ADHM and D-instantons in orbifold 
AdS/CFT duality}

\author{Timothy J.~Hollowood$^{a,c}$ and Valentin V.~Khoze$^b$\\
$^a$Theoretical Division T-8, Los Alamos National Laboratory,
Los Alamos, NM 87545, USA\\
$^b$Department of Physics, University of Durham,
Durham, DH1 3LE, UK\\
$^c$Department of Physics, University of Wales Swansea,
Swansea, SA2 8PP, UK\\
E-mail: {\tt pyth@skye.lanl.gov}, {\tt valya.khoze@durham.ac.uk}} 

\abstract{We consider ADHM instantons in product group gauge theories that
arise from D3-branes
located at points in the orbifold ${\Bbb R}^6/{\Bbb Z}_p$. At finite
$N$ we argue that the ADHM construction and collective coordinate integration
measure can be deduced from
the dynamics of D-instantons in the D3-brane background. For the
large-$N$ conformal field theories of this type, we compute a saddle-point
approximation of the ADHM integration measure and show that it is proportional to the
partition function of D-instantons in the dual $AdS_5\times S^5/{\Bbb
Z}_p$ background, in agreement with the orbifold AdS/CFT 
correspondence. Matching the expected behaviour of D-instantons,
we find that when $S^5/{\Bbb Z}_p$ is smooth 
a saddle-point solution only exists in the sector where the 
instanton charges in each gauge group factor are the same.
However, when $S^5/{\Bbb Z}_p$ is singular, the instanton charges
at large $N$ need not be the same and the space of saddle-point solutions has
a number of distinct branches which represent the possible fractionations of 
D-instantons at the singularity. For the theories with a type 0B dual
the saddle-point solutions manifest two types of D-instantons.}

\keywords{Solitons Monopoles and Instantons, $1/N$ Expansion, Duality
in Gauge Field Theories, Supersymmetry and Duality}

\preprint{{\tt hep-th/9908035}}

\begin{document}

\section{Introduction}

There is a natural relation between D-branes and the ADHM construction
of multi-instanton solutions in gauge theory which arises from considering
a configuration of $k$ D$p$-branes
embedded in $N$ coincident D$(p+4)$-branes. The embedding is described by a
charge $k$ instanton solution in the four-dimensional $\SU(N)$ gauge theory 
associated to the four transverse directions of the
D$p$-branes in the D$(p+4)$-brane world volume
\cite{Witten:1996gx,Douglas:1995bn,Douglas:1996uz,MO-III}. 
This correspondence is valid for arbitrary values of $N$, but in the large-$N$
limit it simplifies dramatically \cite{MO-III} leading to powerful
instanton tests  \cite{lett,MO-III,Green:rev}
of the AdS/CFT duality in $\N=4$ gauge theory
\cite{AGMOO:rev,MGKPW}.
Thus both at finite $N$ and
in the large-$N$ limit there is a very appealing relation between ADHM
and D-instantons. It is this relation in more general theories that we want to
further investigate in this paper. 

To see how the collective coordinate integration measure for ADHM
multi-instantons emerges from D-brane dynamics, consider
$k$ D$(-1)$-branes (D-instantons) embedded in $N$ coincident D3-branes.
The theory on the world volume of $N$ D3-branes is four-dimensional
$\N=4$ supersymmetric $\SU(N)$ gauge theory. The $k$ D-instantons correspond
to `point-like' topological defects identified
with Yang-Mills instantons of charge $k$. The theory on the world volume
of the D-instantons describes the Yang-Mills instanton dynamics.
This is a zero-dimensional matrix theory whose partition function
in the strong coupling $\alpha' \rightarrow 0$ limit gives the 
ADHM collective coordinate 
integration measure (weighted with the instanton action).
This relation holds for arbitrary $N$ and the
$\alpha' \rightarrow 0$ limit is taken in order to decouple the world volume
theory from gravity in the bulk.
This finite-$N$ approach was used in Sec. IV.2 of \cite{MO-III}
to derive directly from D-brane dynamics the 
ADHM measure in $\N=4$ supersymmetric $\SU(N)$ gauge theory. The
resulting expression is in precise agreement with the expression
previously deduced in Ref.~\cite{meas1,meas2} from the field theory
considerations alone.

The matrix model describing the D-instantons inside the
D3-branes can be viewed as a dimensional reduction to zero dimensions
of the effective theory describing D5-branes inside D9-branes.
The theory on the world volume of the $k$ D5-branes 
is pure $\N=(1,1)$ supersymmetric $\U(k)$
gauge theory in six dimensions with $N$, the number of
D9-branes, additional hypermultiplets (and so the
resulting theory actually only has $\N=(1,0)$ supersymmetry).
What is particularly striking
is that the auxiliary degrees-of-freedom $\chi$
introduced in \cite{lett,MO-III} to
bi-linearize the four-fermion interaction in the ADHM instanton
action arise now in a very natural way as the scalars corresponding to
the six-dimensional gauge field \cite{MO-III}
and describe the freedom for the D-instantons to be ejected from the D3-branes. 
This geometrical interpretation of $\chi$ variables deserves a comment. 
One might think that since we have started
with D$5$-branes inside D$9$-branes and then dimensionally reduced
six dimensions common to the both types of branes, we must end up
with D-instantons lying inside the D3-branes, however, this static
reasoning however is na\"\i ve. The six-dimensional gauge field $\chi$
living on the world volume of the D5-branes turns into six scalar `fields'
after the dimensional reduction. The six scalars
$\chi$ specify excitations of the D-instantons transverse
to the D3 world volume. Thus, when $\chi$ are non-zero 
the D-instantons are in fact ejected from the world volume of the D3-branes.
In the large-$N$ limit, it turns out that $\chi$ gains a `VEV' which
constrains its length and an $S^5$ is generated!

Ref.~\cite{MO-III} went on the consider the large-$N$ limit of this
measure. Using a steepest descent approximation the result is very simple to
state. The large-$N$ gauge invariant ADHM measure is proportional to
the partition function of the six-dimensional pure  $\N=(1,1)$ 
supersymmetric $\U(k)$ gauge theory, with no additional matter,
dimensionally reduced to zero dimensions.
Alternatively, this can be described as the 
ten-dimensional $\N=1$ supersymmetric pure $\U(k)$ gauge theory dimensionally
reduced to zero dimensions. This is
precisely what one expects for D-instantons in a flat background with
no D3 branes present, where the $\U(1)\subset\U(k)$ components of the
ten-dimensional gauge field are interpreted as
the position of the charge-$k$ D-instanton in ${\Bbb R}^{10}$. 
The only difference with the
large-$N$ measure is that the $U(1)$ components of the gauge field are now
interpreted as the position of the charge-$k$ D-instanton
in $AdS_5\times S^5$, the near horizon
geometry of the $N\rightarrow\infty$ D3-branes. 

To make this more precise it is convenient to view the gauge-invariant ADHM
integration measure as a product of two factors: the centre-of-mass measure
and the reduced measure. The centre-of-mass measure includes only
the global collective coordinates---the position, the scale-size,
the supersymmetric and superconformal fermion zero modes, and the
$\chi$-coordinates of the multi-instanton configuration
as the whole. The second factor---the reduced measure---includes
integrations over the remaining relative collective coordinates 
(relative positions, relative scale sizes, etc.).
It follows from the analysis of Ref.~\cite{MO-III} that
the reduced gauge-invariant ADHM measure in the large-$N$ limit
equals the non-abelian part of the D-instanton measure in the flat
background, while the centre-of-mass ADHM measure
gives the volume element of $AdS_5\times S^5$.

This concludes the review of the finite-$N$ and large-$N$ relations
between D-instantons and Yang-Mills instantons in $\N=4$ gauge theory.
We are now ready to generalize this picture other theories with $\N<4$. 
An interesting class of generalizations of the AdS/CFT correspondence
\cite{MGKPW,AGMOO:rev}
to four dimensional theories with less than $\N=4$ supersymmetry, follows from considering 
string theory on a background where the $S^5$ is replaced by the
quotient $S^5/\Gamma$, where in general the finite group $\Gamma$ does not
necessarily act freely and so the resulting space has singularities 
\cite{Kachru:1998ys,Lawrence:1998ja,Bershadsky:1998mb}. For simplicity
here we shall only consider the abelian cases $\Gamma={\Bbb Z}_p$, although our
results have an obvious generalization to the non-abelian cases.
The relevant
string theory is either type IIB, or the non-supersymmetric type 0B,
depending upon the action of ${\Bbb Z}_p$ on the fields of the theory.
There is substantial evidence
that the dual gauge theory is simply a certain ${\Bbb Z}_p$-projection of the
$\N=4$ theory, where ${\Bbb Z}_p$ acts both on gauge and on the $\SU(4)_R$ indices of the
various fields \cite{Lawrence:1998ja}. The resulting theories can have
either $\N=0,1$ or 2 supersymmetries. 

Before summarizing our findings, we briefly review the description of
the ${\Bbb Z}_p$-projected gauge theories and their relation with
D-branes on orbifolds.

\subsection{The projected $\N=4$ gauge theories and D-branes on orbifolds}

The theories that we will consider can be defined as projections of
an ${\cal N}=4$ supersymmetric gauge theory with a unitary gauge
group $\U(n)$ \cite{Kachru:1998ys,Lawrence:1998ja,Bershadsky:1998mb}.
The projection is obtained by an action of a finite
group $\Gamma$ which is embedded in both the gauge group and the
$\SU(4)_R$ group of $R$-symmetries. The embeddings $\Gamma\subset\U(n)$
and $\Gamma\subset\SU(4)_R$ are
specified by the decomposition of the adjoint representation of the
gauge group and the $\bf4$ of $\SU(4)_R$, respectively, into the irreducible
representations of $\Gamma$. The projected theory is then defined by
the same action of the ${\cal N}=4$ theory with all the $\Gamma$
non-invariant fields set to zero. 

The construction outlined above works for any such $\Gamma$, however
for simplicity, we shall concentrate on the abelian cases where
$\Gamma={\Bbb Z}_p$. Without loss of generality we can take the 
embedding ${\Bbb Z}_p\subset\U(n)$ to be generated by
\begin{equation}
\sigma_{\{N_q\}}=\begin{pmatrix} 1_{\sst[N_1]\times[N_1]} &&&\\
& e^{2\pi i/p}1_{\sst[N_2]\times[N_2]} & & \\
& &\ddots  & \\
&&& e^{2(p-1)\pi i/p}1_{\sst[N_p]\times[N_p]}\end{pmatrix}\, .
\label{ggamma}\end{equation}
The embedding ${\Bbb Z}_p\subset\SU(4)_R$ can be
defined by a set of four phases, or equivalently four
integers $\{q_1,q_2,q_3,q_4\}$ defined modulo $p$,
subject to the condition
\begin{equation}
q_1+q_2+q_3+q_4=0\ \text{mod}\,p\ .
\label{sumq}\end{equation}
The gauge field $v_n$ is an $\SU(4)_R$ singlet and hence invariant under  
${\Bbb Z}_p\subset\SU(4)_R$. 
On the fermions $\lambda^A$, which transform as a $\bf4$ of
$\SU(4)_R$, the action of ${\Bbb Z}_p\subset\SU(4)_R$ is generated by 
\begin{equation}
\lambda^A\rightarrow e^{2\pi iq_A/p}\lambda^A\ . 
\label{gsu}\end{equation}
The action on the scalar fields which form a ${\bf 6}$ of $\SU(4)_R$
(or a vector of $\SO(6)\subset \SU(4)_R$)
is given most easily by writing them
as an anti-symmetric tensor $A^{AB}$, subject to the
reality condition\footnote{Here $\dagger$ acts only on gauge indices
and not on $\SU(4)_R$ indices. We frequently pass between the
antisymmetric tensor representation $A^{AB}$ and explicit $\SO(6)$ vector
representation $A_a$, via $A^{AB}=(1/\sqrt8)\bar\Sigma_a^{AB}A_a$,
where the coefficients are defined in the Appendix of
\cite{MO-III}. In particular the inner product $A_aB_a\equiv\epsilon_{ABCD}A^{AB}B^{CD}$.}
\begin{equation}
(A^{AB})^\dagger=\tfrac12\epsilon_{ABCD}A^{CD}\,.
\end{equation}
In this basis the action of ${\Bbb Z}_p\subset\SU(4)_R$ is generated by
\begin{equation}
A^{AB}\rightarrow e^{2\pi i(q_A+q_B)/p}A^{AB}\ .
\label{gso}\end{equation}
The ${\Bbb Z}_p$ action on $\SO(6)$ vectors as specified above 
may not give a faithful representation of ${\Bbb Z}_p$. This happens
when all the combinations $q_A+q_B$ are even and when  $p$ itself is even, in
which case only the subgroup ${\Bbb Z}_{p/2}\subset{\Bbb Z}_p$ is faithfully
represented on vectors, a subtlety which will prove to be important.

We are now in a position to implement the ${\Bbb Z}_p$ projection on
the fields. Taking a combination of the gauge transformation
\eqref{ggamma} and the action on $SU(4)_R$ indices \eqref{gsu}:
\begin{equation}
\sigma_{\{N_q\}}v_n\sigma_{\{N_q\}}^{-1}=v_n\ ,\quad
\sigma_{\{N_q\}}\lambda^A\sigma_{\{N_q\}}^{-1}= e^{-2\pi
iq_A/p}\lambda^A\ ,\quad
\sigma_{\{N_q\}}A^{AB}\sigma_{\{N_q\}}^{-1}= e^{-2\pi
i(q_A+q_B)/p}A^{AB}\ .
\label{proj}\end{equation}
Hence the gauge group of the projected theory is
$\U(N_1)\times\cdots\times \U(N_p)$ with $n=\sum_{q=1}^p N_q$. 
An element of the gauge group of the projected theory has the block
diagonal form
\begin{equation}
\begin{pmatrix} U_{\sst[N_1]\times[N_1]}^{(1)} &&&\\
& U_{\sst[N_2]\times[N_2]}^{(2)} & & \\
& &\ddots  & \\
&&& U_{\sst[N_p]\times[N_p]}^{(p)}\end{pmatrix}\, ,
\label{ggroup}\end{equation}
where $U^{(q)}\in \U(N_q)$. The abelian components of the $\U(N_q)$
factors actually decouple in the infra-red and so effectively we can
take the gauge group to be
$\SU(N_1)\times\cdots\times\SU(N_p)$.\footnote{This decoupling is
valid in the context of a four dimensional theory, however, we shall
also use the same kind of projection in lower dimensional theories
where the decoupling does not occur.}
In order to write down the projected fields we introduce a block-form
notation for the $\U(n)$ adjoint-valued fields of the parent
theory. Let $E_{q\,r}$ be the $p\times p$ matrix with a one in
position $(q,r)$ and zeros elsewhere.\footnote{We will always
think of the labels $q,r,q_A$, etc., as being defined modulo $p$.} 
The gauge field of the
projected theory has the block-diagonal form
\begin{equation}
v_n=\sum_{q=1}^pv^{(q)}_n\otimes E_{q\,q}\ ,
\end{equation}
where the block $v^{(q)}_n$ is the gauge field of the
$\U(N_q)$ subgroup of the gauge group. The fermions and scalars have
off-diagonal components, since the scalar and fermion fields in the original theory have
non-trivial transformations under the $R$-symmetry:
\begin{equation}
\lambda^A=\sum_{q=1}^p\lambda^{(q)A}\otimes E_{q\,q+q_A}\ ,\qquad
A^{AB}=\sum_{q=1}^pA^{(q)AB}\otimes E_{q\,q+q_A+q_B}\ .
\label{bferm}\end{equation}

One can see that the spectrum of fields in the theory consists of the
gauge bosons of the product gauge group along with various matter
fields that transform in bi-fundamental representations $(N_q,\bar
N_r)$ of a pair of the group
factors, or adjoint representations of a single $\U(N_q)$ factor.
The amount of supersymmetry depends on the set of integers
$\{q_1,q_2,q_3,q_4\}$. We can take (subject
to \eqref{sumq}):

(i) $\{0,0,q_3,q_4\}$ (without loss of generality we can take
$q_3=-q_4=1$). These models have ${\cal N}=2$ supersymmetry and
${\Bbb Z}_p$ leaves two components of an $\SO(6)$ vector fixed (in
$\SU(4)$ language the two components $A^{12}$ and $A^{34}$). 

(ii) $\{0,q_2,q_3,q_4\}$ with $q_2,q_3,q_4\neq0$ mod $p$. These models have ${\cal
N}=1$ supersymmetry 

(iii) All the other case have $\N=0$ supersymmetry.

These theories describe the low energy dynamics of D3-branes on the
orbifold ${\Bbb R}^4\times{\Bbb R}^6/{\Bbb Z}_p$, where the D3-branes
lie along the ${\Bbb R}^4$ factor and so at a point on the
${\Bbb R}^6/{\Bbb Z}_p$
orbifold \cite{Douglas:1996sw}. One way to analyze this set-up is to
consider the D3-branes
on the covering space ${\Bbb R}^{10}$ using a method of images. 
Each D3-brane will then come
with its images under ${\Bbb Z}_p$. So each collection of $N$ D3-branes at
the same point in ${\Bbb R}^6$ will have $p$ images. This naturally
gives rise to the gauge theory described above but with $N\equiv
N_1=\cdots=N_p$. This is situation where the adjoint of the original
$\U(n)\equiv\U(pN)$ theory decomposes into $N$ copies of the regular
representation of ${\Bbb Z}_p$. However, there is more freedom if the
D3-branes are at the singularity of the orbifold
\cite{Douglas:1996sw}. In that case, there
are no images and consequently one can permit the more general situation
with gauge group $\U(N_1)\times\cdots\times\U(N_p)$. Each $\U(N_q)$ factor
describes $N_q$ `fractional' D3-branes which are confined to move on
the orbifold singularity \cite{Douglas:1996sw,Douglas:1997xg,Diaconescu:1998br}. 
In this case $p$ fractional D3-branes of each of the 
$p$ types can form a genuine D3-brane which can then move away from the
singularity.\footnote{It is interesting to view these fractional
D3-branes in a T-dual set-up \cite{Karch:1998yv} where they correspond
to a segment of D4-brane suspended between two NS5-branes as in \cite{Witten:1997sc}.}

It turns out that it is precisely the theories 
where the D3-branes can roam on the orbifold, i.e.~$N\equiv
N_1=\cdots=N_p$, 
which are relevant for large-$N$ limit and an AdS/CFT
duality, because these theories are, for $\N=1$ and 2, conformal, and,
for $\N=0$, conformal at leading order in $1/N$ \cite{Bershadsky:1998mb}. 
For the supersymmetric theories the large-$N$ dual is conjectured to be 
type IIB superstring theory on $AdS_5\times S^5/{\Bbb Z}_p$
\cite{Kachru:1998ys}. For
$\N=2$ supersymmetry the action of ${\Bbb Z}_p$ fixes an $S^1\subset S^5$ and so the
resulting spacetime has an orbifold singularity and consequently there is no
supergravity description (unless the singularity is blown up in
some way). On the contrary for the $\N=1$ theories there are no singularities and we
expect to have a supergravity description in the appropriate limit.
Notice that in this case there is an important difference between the flat
orbifold ${\Bbb R}^4\times{\Bbb R}^6/{\Bbb Z}_p$ and the near horizon
geometry $AdS_5\times S^5/{\Bbb Z}_p$. The former has an orbifold
singularity at the origin of ${\Bbb R}^6$, whereas there is no such
singularity in the latter since
the radius of the $S^5$ is constant over $AdS_5$
\cite{Kachru:1998ys}. This fact will turn
out to have an important implication for the large-$N$ limit of the
instanton measure.

For the $\N=0$ theories the situation is slightly more complicated because
there are two cases depending on whether ${\Bbb Z}_p\subset
\SO(6)\subset\SU(4)$ or not. In general only the subgroup 
${\Bbb Z}_{p/2}\subset{\Bbb Z}_p$ acts faithfully on $\SO(6)$
vectors. In this case the element whose action on $\SO(6)$ vectors is
not faithfully represented acts on the fields
as $(-1)^F$, where $F$ is the fermion number operator.
When ${\Bbb Z}_p$ does act faithfully then the large-$N$ dual is expected to be
the type IIB superstring on $AdS_5\times S^5/{\Bbb Z}_p$, as for the
supersymmetric cases. However, when only ${\Bbb Z}_{p/2}$ acts
faithfully, the appearance of the factor $(-1)^F$ is a clue that in this case the
dual theory is not the type IIB superstring but rather is the
non-supersymmetric type 0B string.
In particular, when $p=2$, this is precisely the projection of the
$\N=4$ which is thought to be dual to the non-supersymmetric type 0B string
on $AdS_5\times S^5$ \cite{Nekrasov:1999mn,Klebanov:1999ch}. 
In general we expect the dual theory to be the
type 0B string on $AdS_5\times S^5/{\Bbb Z}_{p/2}$.  

\subsection{Summary of results}

As in Ref.~\cite{MO-III} we can investigate the general ${\Bbb
Z}_p$-projected theories at finite $N$
by considering the dynamics of D3-branes and D-instantons in the flat
orbifold background ${\Bbb R}^4\times{\Bbb R}^6/{\Bbb Z}_p$, where the
D3-branes lie along ${\Bbb R}^4$ and hence at a point on the
orbifold ${\Bbb R}^6/{\Bbb Z}_p$. In particular if the D3-branes lie 
at the orbifold
singularity then the resulting gauge theory can have the general product
group structure $\U(N_1)\times\cdots\times\U(N_p)$ with matter fields
transforming in bi-fundamental representations generalizing to a
six-dimensional orbifold the set-up
in Ref.~\cite{Douglas:1996sw}. This theory is the
${\Bbb Z}_p$-projection of the $\N=4$ supersymmetric gauge
theory with gauge group $\U(n)$, $n=\sum_{q=1}^pN_q$, described in
the last section. We will argue that the matrix theory of D-instantons which lie at the
orbifold singularity in the presence of the D3-branes 
is a certain ${\Bbb Z}_p$-projection 
of the $\U(K)$ matrix theory
describing the D-instantons in the $\N=4$ case, as described in more
detail in Sec.~3.2.\footnote{In our
notation $k_q$ is the
instanton charge for the $\U(N_q)$ factor of the gauge
group. We also define $K=k_1+\cdots+k_p$ to be the total instanton
charge.} The discrete group ${\Bbb Z}_p$ is
embedded in the $\U(K)$ group of the matrix theory as $\sigma_{\{k_q\}}$ and in the
$\U(n)$ `flavour' group as $\sigma_{\{N_q\}}$.
The group also acts, as in \eqref{gsu}, on $\SU(4)_R$, which in the matrix 
model is the covering group of the Lorentz group of the 
parent six-dimensional theory. The resulting theory
consequently has symmetry groups $\U(k_1)\times\cdots\times\U(k_p)$
and $\U(N_1)\times\cdots\times\U(N_p)$.
We argue that the partition function of the resulting matrix model in
the decoupling limit $\alpha'\rightarrow0$ is 
precisely the ADHM measure in the gauge theory in the instanton charge
sector ${\cal C}=\{k_1,\ldots,k_p\}$, generalizing the
$\N=4$ case in \cite{MO-III} in an obvious way. In particular, in the
$\N=2$ case the resulting theories arise from the dimensional
reduction of the D3 and
D7-brane configurations in the ${\Bbb R}^4/{\Bbb Z}_p$ orbifold
background described in \cite{Douglas:1996sw}.

We then consider the large-$N$ gauge invariant measure for ADHM
instanton in the conformal theories with gauge
group $\U(N)\times\cdots\times\U(N)$ ($p$-times) and find the following 
results. Firstly, in (i) and (ii) below, for the theories
whose dual is the type IIB theory (i.e.~all the supersymmetric
theories and the $\N=0$ theories where ${\Bbb Z}_p$ acts faithfully on
$\SO(6)$ vectors):

(i) When ${\Bbb Z}_p$ acts freely on $\SO(6)$ vectors, a solution of the
large-$N$ saddle-point equations for the instanton measure only exists
in the sector where all the instanton charges are the same.
In other words when $S^5/{\Bbb Z}_p$ is smooth in the dual theory, 
only the ${\cal C}=\{k,\ldots,k\}$ charge sectors contribute at
leading order in $1/N$. The saddle-point solution describes $k$
point-like objects, the D-instantons of the string theory, moving in 
$AdS_5\times S^5/{\Bbb Z}_p$.
The large-$N$ instanton measure has the form of the partition 
function of the ${\Bbb Z}_p$-projected $\N=1$
supersymmetric $\U(K)\equiv\U(pk)$ gauge theory in ten dimensions, 
dimensionally reduced to zero dimensions. This theory has gauge group
$\U(k)\times\cdots\times\U(k)$. Equivalently we may describe it as 
the ${\Bbb Z}_p$-projection of the $\N=(1,1)$ pure
six-dimensional $\U(K)$ gauge theory dimensionally reduced to zero dimensions.
The saddle-point solution corresponds to the coulomb branch of
the matrix theory where the $\U(k)\times\cdots\times\U(k)$ gauge
symmetry is generically broken to $\U(1)^k_{\sst\rm
diag}$.\footnote{This is the maximal abelian subgroup of the diagonal
subgroup $\U(k)_{\rm diag}\subset\U(k)\times\cdots\times\U(k)$.}

(ii) When ${\Bbb Z}_p$ does not act freely on $\SO(6)$ vectors a
solution of the saddle-point equations exists in the general charge
sector where the instanton numbers can be different
${\cal C}=\{k_1,\ldots,k_p\}$. In this case, the saddle-point
solutions have multiple branches labelled by $\tilde k=0,\ldots,{\rm
min}(k_q)$. The branches describe $\tilde k$ D-instantons moving in 
$AdS_5\times S^5/{\Bbb Z}_p$ and $k_q-\tilde k$ fractional
D-instantons of type $q$ moving in $AdS_5\times(S^5/{\Bbb Z}_p)_{\rm
sing}$. (Here $(S^5/{\Bbb Z}_p)_{\rm
sing}$ is the subspace of $S^5$ fixed by the action of ${\Bbb Z}_p$.) 
The large-$N$ instanton measure is then described by the partition
function of the ${\Bbb Z}_p$-projected 
$\N=1$ ten-dimensional $\U(K)$ gauge theory dimensionally reduced to
zero dimensions, as in (i), but with the more general 
${\Bbb Z}_p$-projection which permits the gauge group
$\U(k_1)\times\cdots\times\U(k_p)$ (or as in (i) 
equivalently the ${\Bbb Z}_p$-projection of the $\N=(1,1)$ pure
six-dimensional $\U(K)$ gauge theory dimensionally reduced to zero dimensions).
The saddle-point solutions correspond to a series of coulomb branches of the
matrix theory where the symmetry $\U(k_1)\times\cdots\times\U(k_p)$ is generically
broken to $\U(1)_{\sst\rm diag}^{\tilde k}
\times\U(1)^{K-p\tilde k}$. When $\tilde k=k_1=\cdots=k_p$ this is
identical to the coulomb branch in (i) above.

(iii) In the non-supersymmetric theories with a type 0B dual where only  
${\Bbb Z}_{p/2}\subset{\Bbb Z}_p$ acts faithfully on $\SO(6)$ vectors, the
solution of the saddle-point equations is slightly more complicated. 
In this case, we split the instantons into two sets with charges
\begin{equation}
{\cal C}_+=\{k_2,k_4,\ldots,k_p\}\ ,\qquad
{\cal C}_-=\{k_1,k_3,\ldots,k_{p-1}\}\ .
\end{equation}
~From the point-of-view of the saddle-point analysis the two sets of
instantons are completely decoupled.\footnote{The point is that the
two sets only communicate through fermionic variables which play no
r\^ole in the saddle-point analysis.} For each set ${\cal C}_\pm$ 
separately the solutions of the
saddle-point equations is analogous to (i) and (ii) above. So if
${\Bbb Z}_{p/2}$ acts freely on $S^5$ then a saddle-point solution
only exists if each of the instanton charges in each sets ${\cal
C}_\pm$ is the same: ${\cal C}_\pm=\{k_\pm,\ldots,k_\pm\}$. In this case the
solution can be interpreted as describing two kinds of D-instantons in
$AdS_5\times S^5/{\Bbb Z}_{p/2}$. On the contrary, if ${\Bbb Z}_{p/2}$
does note act freely on $S^5$, then saddle-point solutions exist in
the general $\{k_1,\ldots,k_p\}$ charge sectors. As in the type IIB
cases the solutions now exhibit multiple branches describing the fractionation
of the two kinds of D-instantons.

Beginning with the type IIB cases, 
our results are exactly what one would expect from generalizing the
result of the same analysis in the $\N=4$ theory \cite{MO-III}. In other words we
find that the large-$N$ ADHM instanton measure is identical to
the D-instanton partition function on the
flat orbifold ${\Bbb R}^4\times{\Bbb R}^6/{\Bbb Z}_p$. This is to be
expected: as in the $\N=4$ theory the partition function for
D-instantons on $AdS_5\times S^5$ is identical to that in flat space. 
However, there is an important restriction in the cases when
$S^5/{\Bbb Z}_p$ is smooth. In the corresponding flat orbifold,
describing the finite-$N$ solution, the
D-instantons could lie at the singularity and one would have
contributions from the general
charge sector $\{k_1,\ldots,k_p\}$ corresponding to 
fractional D-instantons. However the near horizon geometry $AdS_5\times S^5/{\Bbb
Z}_p$ is smooth and so we expect only the charge sector
$\{k,\ldots,k\}$ can contribute at large $N$ and this is precisely
what we find from our
saddle-point analysis for case (i). When $S^5/{\Bbb Z}_p$ has
a singularity there is no such constraint on the charge sector as we
find from the saddle-point analysis in case (ii). In this case the
space of solutions has a number of different branches which represents
the possible fractionations of D-instantons at the singularity
$(S^5/{\Bbb Z}_p)_{\rm sing}$.

For the theories whose duals are the type 0B string theory the
situation is essentially doubled up: in this case the instantons in
the even/odd groups in the product $\U(N)\times\cdots\times\U(N)$ are
independently interpreted as in the last paragraph with 
D-instantons in the string theory. It is interesting that this is exactly
what is expected from the type 0B string theory since the
number of Ramond-Ramond fields is doubled up compared with the type
IIB theory. This means that each D-brane comes in two varieties:
`magnetic' and `electric'. It is pleasing that this feature also
emerges as the outcome of a saddle-point analysis of large-$N$ gauge theory instantons. 

\section{The ADHM Construction for Product Groups}

In this section we consider the construction of multi-instanton solutions
for the projected $\N=4$ theory. The construction is a very obvious
generalization of the ADHM formalism set out in \cite{MO-III}. The new feature
is that some of the matter fields transform in bi-fundamental
representations of a pair of the gauge group factors.
Fortunately this problem was considered in the early instanton
literature \cite{Corrigan:1979xi} as part of a program to construct solutions
for fields transforming in arbitrary representations of a gauge group
in the back-ground of an ADHM instanton. 
The resulting formalism is such an obvious generalization of that for
an adjoint-valued field that we will keep our discussion brief.
Note that as far as the instanton solutions are concerned there is no
difference between $\U(N_q)$ and $\SU(N_q)$ since the instanton is
embedded in the non-abelian part of the gauge group.

\subsection{The gauge fields}

First of all consider the gauge fields. To leading order in the
coupling one simply ignores the matter fields and considers the gauge
fields in isolation. At this order, there are no couplings between the
different $\SU(N_q)$ group factors; in order words, one constructs a general
instanton solution by embedding instanton solutions with charges
$k_q$, $q=1,\ldots,p$, in each of the $\SU(N_q)$ factors. The ADHM
construction then involves the familiar objects $U^{(q)}$, $\bar
U^{(q)}$, $\Delta^{(q)}$,
$a^{(q)}$, $b^{(q)}$, $f^{(q)}$, $a^{\prime(q)}_n$, $w_\aD^{(q)}$,
etc., for each of the
groups.\footnote{We will assume that the reader is familiar with
Sec.~II of \cite{MO-III}.}
As with the gauge fields of the theory, it is convenient to
assemble the ADHM parameters for the product group 
into $p\times p$ block diagonal matrix where each block pertains to a
single $\SU(N_q)$ factor. So, for example
\begin{equation}
a=\sum_{q=1}^pa^{(q)}\otimes E_{q\,q}\,.
\label{V30}
\end{equation}
The matrix $a^{(q)}$ is then the ADHM matrix of the $q^{\rm th}$ gauge
group factor, and hence is $(N_q+2k_q)\times2k_q$ dimensional. 

The ADHM variables, modulo the ADHM constraints, 
parameterize the moduli space of the instanton solution,
up to an auxiliary symmetry
\begin{equation}
\U(k_1)\times\cdots\times\U(k_p)\,,
\label{agp}\end{equation}
where each factor $\U(k_q)$ acts on the variables of the $q^{\rm th}$
block in the way described in \cite{MO-III}.

\subsection{The fermion fields}

Up till now we have simply used $p$ independent copies of the ADHM construction
one for each of the $\SU(N_q)$ group factors. This is obvious because the
gauge fields of each $\SU(N_q)$ factor of the instanton solution are
completely decoupled. For any matter field which is in an adjoint
representation of the gauge group, an eventuality that will occur if
some $q_A=0$, for the fermions, or some $q_A+q_B=0$, for the scalars,
the construction of the solution is exactly as \cite{MO-III}.
The matter fields, however, can communicate
between the different factors since some of them are in bi-fundamental
representations. 

Let us consider the general problem of a fermion $\lambda_\alpha$ 
transforming in the bi-fundamental
$(N_q,\bar N_r)$ of $\SU(N_q)\times \SU(N_r)$. The equation we have to
solve is the Dirac equation  $\Dbarslash^{\aD\alpha}\lambda_\alpha=0$ 
for $\lambda_\alpha$ in the background of an
instanton solution with charge $k_q$ and $k_r$ in each of the two
group factors, respectively. The solution 
generalizes that for an adjoint fermion \cite{Corrigan:1979xi} and has the form
\begin{equation}
\lambda_\alpha=\bar U^{(q)}{\cal M}^{(q\,r)}f^{(r)}\bar b^{(r)}_\alpha
U^{(r)}-\bar U^{(q)} 
b^{(q)}_\alpha f^{(q)}\bar{\cal M}^{(q\,r)}U^{(r)},
\label{fle} \end{equation}
where $U^{(q)}$ and $U^{(r)}$, etc., refer to objects from the ADHM
construction for the groups $\SU(N_q)$ and $\SU(N_r)$, respectively. The `off-diagonal'
objects ${\cal M}^{(q\,r)}$ and $\bar{\cal M}^{(q\,r)}$
are constant Grassmann matrices of dimension $(N_q+2k_q)\times k_r$
and $k_q\times(N_r+2k_r)$, respectively. The Dirac equation is then
satisfied by virtue of the fermionic ADHM constraints 
\begin{equation}
\bar\Delta^{(q)}_\aD{\cal M}^{(q\,r)}+\bar{\cal M}^{(q\,r)}\Delta^{(r)}_\aD=0\,.
\label{flc} \end{equation}
Equations \eqref{fle},\eqref{flc} correctly reproduce the adjoint fermion zero
mode construction \cite{Corrigan:1979xi,DKMS,MO-III} when $N_q=N_r$ and $k_q=k_r$,
while when $k_r=0$ the $\SU(N_q)$ fundamental fermion zero mode
construction is recovered
\cite{Corrigan:1979xi,DKMS}. In the latter case
$\Delta^{(r)}_\aD$ and ${\cal M}^{(q\,r)}$ are trivial, and 
the collective coordinates $\bar{\cal M}^{(q\,r)}$ 
are unconstrained, just as required.

It is now rather easy to see how to generalize the construction when
we have a collection of fermion fields that transform in a set of
adjoint and bi-fundamental representations of the product group as in \eqref{bferm}.
The trick is to write the fermionic collective coordinates in terms of
the block-form matrices already introduced in \eqref{bferm}. To this end we
introduce  ${\cal M}^A$ and $\bar{\cal M}^A$ that are four
$(n+2K)\times K$ and $K\times(n+2K)$
matrices of constant Grassmann numbers with non-zero elements in particular blocks:
\begin{equation}
{\cal M}^A=\sum_{q=1}^p{\cal M}^{(q)A}\otimes E_{q\,q+q_A}\,,
\label{bfm}\end{equation}
where ${\cal M}^{(q)A}\equiv{\cal M}^{(q\,q+q_A)A}$ and
$\bar{\cal M}^{(q)A}\equiv\bar{\cal M}^{(q\,q+q_A)A}$. 
The ADHM constraints for all the
fermions can therefore be written in a single equation as
\begin{equation}
\bar\Delta_\aD{\cal M}^A+\bar{\cal M}^A\Delta_\aD=0\,.
\label{adhmfc}\end{equation}

\subsection{The scalar fields}

In this section, we consider the ADHM formalism for the scalar fields.
The construction for the scalar fields in the background of an
instanton solution follows the same pattern as in the ${\cal N}=2$ theories
\cite{DKMS} and in the ${\cal N}=4$ theories
\cite{MO-III}. 

To leading order, the scalar fields satisfy the covariant Klein-Gordon equation,
with a source term bi-linear in fermions coming from the Yukawa
interactions. Fortunately, the most obvious generalization of the
solution in the ${\cal N}=4$ theory provides the solution
in the projected theories. As in the ${\cal N}=4$ theory it is convenient to
imagine
a set of collective coordinates for the scalar fields, even though no
actual moduli exist. In our theories, these pseudo collective
coordinates take the block form
\begin{equation}
\A^{AB}\ =\ \sum_{q=1}^p\A^{(q)AB}\otimes E_{q\,q+q_A+q_B}
\end{equation}
where $\A^{(q)AB}$ is a $k_q\times k_{q+q_A+q_B}$ matrix. The pseudo
collective coordinates are then eliminated---at some later stage---by
the algebraic equations
\begin{equation}\bigL\cdot\A^{AB}\ =\ \Lambda^{AB}\ .
\label{thirtysomething}\end{equation}
Here the right-hand side involves a bilinear in the fermionic
collective coordinates
\def\Lambdabar{\bar\Lambda}
\begin{equation}\Lambda^{AB}\ =\ {1\over2\sqrtwo}\,
\big(\,\Mbar^A\M^B -\Mbar^B\M^A \,\big)
\label{newmatdef}\end{equation}
and reflects the Yukawa source term. In \eqref{newmatdef},
the linear operator $\BL$ on $K\times K$ matrices is defined by
\begin{equation}
\bigL\cdot\Omega\ =\ 
\hf\{\,\Omega\,,\,W^0\,\}\,+\,[\,a_n'\,,\,[\,a_n'\,,\,\Omega\,]\,]\ ,
\label{bigLreally}\end{equation}
where $K\times K$ hermitian matrix $W^0$ has the
block-diagonal components
\begin{equation}W^{(q)0}\ =\ \wbar^{(q)\aD} \  
w^{(q)}_\aD\,, 
\label{Wdef}\end{equation}
For later use, we define the components of $\BL$ with respect to
the block-form basis:
\begin{equation}
\BL\cdot\sum_{q=1}^p\Omega^{(q)}\otimes E_{q\,q+r}\equiv
\sum_{q=1}^p\BL^{(q\,q+r)}\cdot\Omega^{(q)}\otimes E_{q\,q+r}
\end{equation}
from which we deduce that on $k_q\times k_r$ matrices $\Phi$
\begin{equation}
\BL^{(q\,r)}\cdot\Phi=
(\hf W^{(q)0}+a^{\prime(q)}_na_n^{\prime(q)})\Phi+
\Phi(\hf W^{(r)0}+a_n^{\prime(r)}a_n^{\prime(r)})-
2a^{\prime(q)}_n\Phi a_n^{\prime(r)}\,.
\label{deflt}\end{equation}

\subsection{The multi-instanton action}

When the expressions for the gauge, fermions and scalar fields 
are substituted into the action the result is not simply a constant,
but contains an interaction term which depends on the 
fermionic collective coordinates:
\begin{equation}
S_{\rm inst}\ =\ {8\pi^2K\over g^2}\ -iK\theta\ +\ S_{\rm quad}\ ,
\label{Skinstdef}\end{equation}
where $S_{\rm quad}$ is a fermion quadrilinear interaction
\begin{equation}S_{\rm quad}\ =\ 
{\pi^2\over g^2}\,\epsilon_{ABCD}\,{\rm tr}_K\,\Lambda^{AB}
{\cal A}^{CD}\
=\ {\pi^2\over
g^2}\,\epsilon_{ABCD}\,{\rm tr}_K\,\Lambda^{AB}
\bigL^{-1}\Lambda^{CD}\  .  
\label{Skquadef}\end{equation}
The expression for $S_{\rm quad}$ is derived in a completely
analogous way to the ${\cal N}=4$ theory \cite{MO-III}.

This is somewhat of a surprise: one might have expected the action to be a constant if
we have a solution of the Euler-Lagrange equations. As explained in
\cite{MO-III}, the ADHM expressions for the gauge, fermion and scalar
fields are not, in
fact an exact solution to the Euler-Lagrange equations when all the fermion
modes are `turned-on'. The fact is that although the fermion zero modes
are Dirac zero-modes, the majority of them are lifted at tree level by the Yukawa
interactions with the scalars. The philosophy that we adopt, and
explained at length in \cite{MO-III}, is that retaining collective
coordinates for the lifted modes provide a convenient way of including the
perturbative (tree level) effects of the modes.
In this point-of-view, the only exact fermion zero modes, not lifted
by \eqref{Skquadef}, are the supersymmetric and superconformal zero
modes. These modes are defined in terms of $\M^A$ and
$\bar\M^A$, as in the $\N=4$ case \cite{MO-III}, but only for $q_A=0$, giving 
8, 4 and 0 modes, for ${\cal N}=2$, 1 and 0,
respectively.

\section{The Collective Coordinate Measure}

In this section we write down the integration measure on the space of ADHM
collective coordinates and then show how for $N_q\geq2k_q$ it can be reduced
to a measure on the set of gauge invariant
variables. Then in this new set of variables, the ADHM constraints 
will be explicitly resolved. Note that the restriction $N_q\geq2k_q$ 
is certainly consistent with the large-$N$ limit.

\subsection{The flat measure}

In order to calculate physical quantities we need to know how to
integrate on the space of ADHM variables. This is the measure induced from
the full functional integral of the field theory. Thankfully, 
repeating the argument of Refs.~\cite{meas1,meas2,MO-III}
it turns out that the
measure is remarkably simple when written in terms of the complete
set of bosonic and fermionic
ADHM variables: it is just the flat measure for all
the variables with all algebraic constraints imposed via explicit delta
functions. 
In order to define the physical measure, we must divide by
the volume of the auxiliary group \eqref{agp}:
\begin{equation}\begin{split}
&\int d\mu_{\rm phys}^{\{k_q\}}\ \sim\ 
{1\over\prod_{q=1}^p{\rm Vol}\,U(k_q)}
\int da' \
d\wbar \ dw  
\prod_{A=1,2,3,4}
\ d\M^{\prime A}  
\ d\mubar^A \ d\mu^A\,\prod_{B=2,3,4}
\ d\A^{1B}\\ &\times
\prod_{c=1,2,3} 
\delta^{[0]}\big(\trtwo\, \tau^c \abar a\big)
\prod_{A=1,2,3,4}
\prod_{\aD=1,2}\delta^{[q_A]}\big(\Mbar^A a_\aD + \abar_\aD \M^A
\big)\,\prod_{B=2,3,4}
\delta^{[q_1+q_B]}\big(\bigL\cdot\A^{1B} -
\Lambda^{1B}\big)\ .
\label{dmudef}\end{split}\end{equation}
The $\sim$ above indicates that in contrast with our previous work
\cite{meas1,meas2,MO-III} we are not going to keep track of the
overall normalization of the measure.
In the above, both the integrals and the delta functions are defined
with respect to a particular basis of matrices. A given $K\times K$ matrix
quantity with block-form $M=\sum_{q=1}^p M^{(q)}\otimes E_{q\,q+r}$,
where $M^{(q)}$ is a $k_q \times k_{q+r}$ matrix, can be expanded in the
basis of $K\times K$ matrices $T^{(r)}_a$:
\begin{equation}
M=\sum_{a=1}^{n_r}M_aT^{(r)}_a\,,
\end{equation}
where $n_r=\sum_{q=1}^pk_qk_{q+r}$ and the basis
$T^{(r)}_a$, with $T^{(r)\dagger}_a=T^{(-r)}_a$, is normalized by 
${\rm tr}_K\,T^{(r)}_aT^{(-s)}_b=\delta^{rs}\delta_{ab}$. The delta
functions in \eqref{dmudef} are defined as
\begin{equation}
\delta^{[r]}(M)=\prod_{a=1}^{n_r}
\delta\big({\rm tr}_K\,T^{(-r)}_aM\big)\ .
\end{equation}
The pseudo collective coordinates for the scalar fields $\A^{AB}$ can be
explicitly integrated out:
\begin{equation}\begin{split}
\int\prod_{B=2,3,4} d \A^{1B} \
&\prod_{B=2,3,4}
\delta^{[q_1+q_B]}\big(\bigL\cdot\A^{1B} - \Lambda^{1B}\big)\\
&=\prod_{q=1}^p\prod_{B=2,3,4}\big({\det}\,\bigL^{(q\,q+q_1+q_B)}\big)^{-1} 
=\prod_{q=1}^p\prod_{A<B}\big({\det}\,\bigL^{(q\,q+q_A+q_B)}\big)^{-1/2}\ .
\label{trivint}\end{split}\end{equation}
The last equality follows from \eqref{sumq} and the fact that 
${\det}\,\bigL^{(q\,r)} ={\det}\,\bigL^{(r\,q)}$.
 
The measure \eqref{dmudef} must be augmented with the instanton action,
$\exp\,-S_{\rm inst}$
and, as we discuss in Sec.~3.3, the ratio of the fluctuation
determinants in the ${\cal N}=0$ cases. Following the ${\cal N}=4$ 
example, it is convenient to bi-linearize the fermion quadrilinear
interaction by introducing a set of auxiliary bosonic `collective
coordinates' $\chi_{AB}$:
\begin{equation}\begin{split}
\prod_{q=1}^p\prod_{A<B}&\big(\det\,\BL^{(q\,q+q_A+q_B)}\big)^{-1/2}\exp\,-S_{\rm
 quad}\\
 &\sim\ \int
d\chi\,\exp\big[-{\rm tr}_K\,\chi_a\BL
\chi_a+4\pi ig^{-1}{\rm
tr}_K\,\chi_{AB}\Lambda^{AB}\big]\ . 
\label{E53}\end{split}\end{equation}
Notice that this transformation absorbs the determinant factors appearing 
in \eqref{trivint}. The auxiliary variables $\chi_{AB}$ form an antisymmetric
pseudo real tensor of $\SO(6)$ whose elements are $K\times K$ matrices
 subject to
\begin{equation}
\tfrac12\epsilon^{ABCD}\chi_{CD}=\chi_{AB}^\dagger\ ,
\label{E54}\end{equation}
where $\dagger$ acts only on instanton indices. The variables $\chi_{AB}$ can be
written as an explicit $\SO(6)_R$ vector $\chi_a$,
$a=1,\ldots, 6$, by using the
coefficients $\Sigma^a_{AB}$ defined in the Appendix of \cite{MO-III}:
\begin{equation}
\chi_{AB}={1\over\sqrt8}\Sigma^a_{AB}\chi_a\, .
\label{E54.1}\end{equation}
The matrices $\chi_{AB}$ have the following block structure
\begin{equation}
\chi_{AB}=\sum_{q=1}^p\chi_{AB}^{(q)}\otimes E_{q\,q-q_A-q_B}\,.
\label{V31}
\end{equation}

\subsection{The ADHM Measure and D-branes}

In this section, we argue that the measure that we have constructed in
terms of the bosonic variables $\{a'_n,\chi_{AB},w_\aD,\bar
w^\aD\}$ and the fermionic variables $\{\M^{\prime A}_\alpha,\mu^A,
\bar\mu^A\}$, can be derived by considering the dynamics of
D-instantons in the background of D3-branes. 

Let us briefly re-cap the same story 
in the $\N=4$ theory
described in Sec.~IV.2 of Ref.~\cite{MO-III}. 
In order to describe the D$(-1)$-D3
system one starts with the $D5-D9$ system and then dimensionally
reduces. We are interested in the world-volume theory of the D5-branes
which on dimensional reduction will describe the matrix model of the
D-instantons. The six-dimensional theory describing the $K$ D5-branes
consists of a $\U(K)$ vector multiplet of $\N=(1,1)$ supersymmetry
with fields $\{\chi_a,\M^{\prime A}_\alpha,\lambda^\aD_A,a'_n,D^c\}$ 
and $n$ fundamental hypermultiplets of $\N=(1,0)$ supersymmetry with fields 
$\{w_\aD,\bar w^\aD,\mu^A,\bar\mu^A\}$.\footnote{The indices are defined as
follows: $a=1,\ldots,6$ and $A=1,\ldots,4$ are spacetime $\SU(4)$
vector and spinor indices; $\alpha=1,2$ and $\aD=1,2$ are spinor
indices of the $\SU(2)_L\times\SU(2)_R$ $R$-symmetry group of the theory;
$n=1,\ldots,4$ is a vector index of
$\SO(4)\subset\SU(2)_L\times\SU(2)_R$ and finally $c=1,2,3$ labels the
adjoint representation of $\SU(2)_R$.}

In Ref.~\cite{MO-III} we showed that the partition function of the
matrix theory resulting from the dimensional reduction to zero
dimensions of the six-dimensional theory whose field content is described above
in the strong coupling
limit (corresponding to $\alpha'\rightarrow0$) is precisely the ADHM
$k$-instanton measure (weighted with the action). The relation between the
fields of the six-dimensional theory and the ADHM construction is
manifest, except for the bosonic variables $D^c$ and 
the fermionic variables $\lambda^\aD_A$. These fields may be eliminated by their
equations-of-motion and they simply enforce the bosonic and fermionic
ADHM constraints by producing the explicit delta functions as in our
ADHM measure \eqref{dmudef}. 

We now describe how this relation extends to the projected
theories. The idea is a simple extension of the $\N=4$ case to the case
where the
D-instantons and D3-branes move on the orbifold spacetime ${\Bbb
R}^4\times{\Bbb R}^6/{\Bbb Z}_p$. The D3-branes lie along ${\Bbb
R}^4$ and at the singularity of the orbifold. The world-volume
theory of the D3-branes is the ${\Bbb Z}_p$-projected $\U(n)$ gauge
theory as described in Sec.~1.2. Now we want to consider the matrix
theory of $K$ D-instantons moving in the D3-brane world volume. Not
surprisingly the resulting theory will be identical to the $\N=4$ case, but with a
${\Bbb Z}_p$ projection. It is easy to see how the projection must act
on the matrices. Firstly it is embedded in the $\U(K)$ symmetry (the
remnant of the six-dimensional gauge symmetry) as $\sigma_{\{k_q\}}$
and in the $\U(n)$ symmetry (the remnant of the flavour symmetry of
the six-dimensional theory) as $\sigma_{\{N_q\}}$. Finally it acts on
$SU(4)$ spacetime indices of the six-dimensional parent theory as in
\eqref{gsu}. Explicitly on the bosonic ADHM variables
\begin{equation}\begin{split}
\sigma_{\{k_q\}}a'_n\sigma_{\{k_q\}}^{-1}=a'_n\ ,&\quad
\sigma_{\{N_q\}}w_\aD\sigma_{\{k_q\}}^{-1}=w_\aD\ ,\\
\sigma_{\{k_q\}}\bar w^\aD\sigma_{\{N_q\}}^{-1}=\bar w^\aD\ ,&\quad
\sigma_{\{k_q\}}\chi_{AB}\sigma_{\{k_q\}}^{-1}=e^{2\pi
i(q_A+q_B)/p}\chi_{AB}\ ,
\end{split}\end{equation}
and on the fermionic ADHM variables
\begin{equation}
\sigma_{\{k_q\}}\M^{\prime A}_\alpha\sigma_{\{k_q\}}^{-1}=e^{-2\pi
iq_A/p}\M^{\prime A}_\alpha\ ,\quad
\sigma_{\{N_q\}}\mu^A\sigma_{\{k_q\}}^{-1}=e^{-2\pi iq_A/p}\mu^A\ ,\quad
\sigma_{\{k_q\}}\bar\mu^A\sigma_{\{N_q\}}^{-1}=e^{-2\pi iq_A/p}\bar\mu^A\ .
\end{equation}
Finally on the Lagrange multipliers for the ADHM constraints
\begin{equation}
\sigma_{\{k_q\}}D^c\sigma_{\{k_q\}}^{-1}=D^c\ ,\quad
\sigma_{\{k_q\}}\lambda^\aD_A\sigma_{\{k_q\}}^{-1}=e^{2\pi
iq_A/p}\lambda^\aD_A\ .
\end{equation}
The resulting theory has the expected symmetries:
$\U(k_1)\times\cdots\times\U(k_p)$, the auxiliary symmetry of the ADHM
construction \eqref{agp}, and $\U(N_1)\times\cdots\times\U(N_p)$, the
gauge symmetry of the original ${\Bbb Z}_p$-projected gauge theory. 
Notice that, as in the $\N=4$ case, the variables $\chi_{AB}$
describe the freedom for the D-instantons to be ejected from the
D3-branes in the orbifold directions, the six directions orthogonal to the D3-brane
world-volume. 

\subsection{Fluctuation determinants for ${\cal N}=0$}

The ADHM collective coordinate measure that we have constructed 
in the non-supersymmetric cases needs to be supplemented with the
determinants of the fluctuations of the various fields, gauge,
fermion, scalar and ghosts, around the instanton solution. Fortunately, in a
supersymmetric theory, it was shown by D'Adda and Di Vecchia
\cite{D'Adda:1978ur}, that the
determinants all cancel in a self-dual, i.e.~instanton
background. The purpose of this section is to consider the
determinants in the $\N=0$ theories 
and show that they will not affect the large-$N$ saddle-point
analysis undertaken in Sec.~4. 
This is not surprising as the theories we consider have
the vanishing beta-function in the large-$N$ limit. 

Let us define $\Delta^{(v,f,s)}_{\BR}$ to be the appropriate
generalized Laplacian operators that govern the fluctuations in the
vector, Weyl fermion and complex scalar fields, respectively, 
around the instanton solution, in
the representation ${\BR}$ of the gauge group.\footnote{For the
vector fields, of course, only the adjoint representation is relevant.} 
For each of these operators we will define a suitably regularized
determinant:
\begin{equation}
\det\,\Delta^{(v,f,s)}_{\BR}=\exp-\Gamma^{(v,f,s)}_\BR\ .
\end{equation}
The details of the regularization procedure will not be relevant for
our purposes. The fluctuations for all the fields will contribute 
\begin{equation}
{\cal F}=\exp\big(\Gamma^{(v)}_{\bf adj}-\Gamma^{(f)}_{\BR_f}
+\Gamma^{(s)}_{\BR_s}\big)
\label{fcont}\end{equation}
to the classical measure. In the above $\BR_f$ and $\BR_s$ are the
representations of the gauge group of the Weyl fermions and complex scalar
fields, respectively.

D'Adda and Di Vecchia \cite{D'Adda:1978ur} proved in an instanton background
\begin{equation}
\Gamma^{(v)}_{\bf adj}=\Gamma^{(s)}_{\bf adj},\qquad
\Gamma^{(f)}_{\BR}=\Gamma^{(s)}_{\BR}\ 
\end{equation}
(in the vector case the ghost contribution is included)
and consequently the contribution \eqref{fcont} is
\begin{equation}
{\cal F}=\exp\big(\Gamma^{(s)}_{\bf
adj}-\Gamma^{(s)}_{\BR_f}+\Gamma^{(s)}_{\BR_s}\big)\ .
\end{equation}
In a supersymmetric theory there is a Weyl fermion superpartner
to the gauge field---the gluino---and a Weyl fermionic superpartner
to each complex scalar field; consequently $\BR_f={\bf adj}+\BR_s$ and
the determinants all cancel, i.e.~${\cal F}=1$. 

The fluctuations will only contribute to measure in the ${\cal
N}=0$ cases where in order to evaluate the contribution we need the
expression for $\Gamma^{(s)}_{\BR}$, where $\BR$ is either an adjoint
representation of $\SU(N_q)$ or a bi-fundamental representation
$(\BN_q,\bar\BN_r)$ of $\SU(N_q)\times \SU(N_r)$. Fortunately, expressions for
these determinants were calculated some time ago \cite{Jack:1980rn}. For a
bi-fundamental representation
\begin{equation}
\Gamma^{(s)}_{(\BN_q,\bar\BN_r)}=N_q\Gamma^{(s)}_{\BN_q}+N_r\Gamma^{(s)}_{\BN_r}
-\log\det\,\BL^{(q\,r)}+{1\over16\pi^2}\int d^4x\log\det
f^{(q)}\ \square^2\log\det f^{(r)}\,.
\label{gambf}\end{equation}
In the above, $f^{(q)}$, is the $k_q\times k_q$ ADHM matrix of 
the $\SU(N_q)$ factor. The operator 
$\BL^{(q\,r)}$ is defined in \eqref{deflt}. 
The result for the adjoint representation of $\SU(N_q)$
follows by a restriction \cite{Jack:1980rn}:
\begin{equation}
\Gamma^{(s)}_{{\bf adj}_q}=2N_q\Gamma^{(s)}_{\BN_q}
-\log\det\,\BL^{(q\,q)}+{1\over16\pi^2}\int d^4x\log\det
f^{(q)}\ \square^2\log\det f^{(q)}\,.
\label{gamadj}\end{equation}
Notice that in both cases \eqref{gambf} and \eqref{gamadj}, the only
$N_q$ dependence comes from the first term only, involving the
determinant of the $\BN_q$ representation.
It is fortunate that the cumbersome regularization dependent
expression for $\Gamma^{(s)}_{\BN_q}$
will not required (see
\cite{Berg:1979ku,Corrigan:1979di,Osborn:1981yf} for details).

To get the fluctuation contribution to the measure, we need to
consider the product over the various fields. Although, the final result
is non-vanishing for the non-supersymmetric theories, significantly
the $N_q$ dependent and regularization dependent parts of \eqref{gambf}
and \eqref{gamadj} cancel between the
bosons and fermions to leave the following---$N_q$ independent--contribution to the measure:
\begin{equation}
{\cal F}=\prod_{q=1}^p\Big\{(F^{(q\,q)})^{-1}\prod_{A<B}
(F^{(q\,q+q_A+q_B)})^{-1/2}\prod_AF^{(q\,q+q_A)}\Big\}\,,
\end{equation}
where
\begin{equation}
F^{(q\,r)}=\det\,\BL^{(q\,r)}\exp-{1\over16\pi^2}\int d^4x\log\det
f^{(q)}\ \square^2\log\det f^{(r)}\ .
\end{equation}

For the supersymmetric theories, since $q_1=0$, we have
\begin{equation}
\prod_AF^{(q\,q+q_A)}=F^{(q\,q)}\prod_{A<B}(F^{(q\,q+q_A+q_B)})^{1/2}\,,
\end{equation}
and so ${\cal F}=1$, as expected from the general analysis of D'Adda
and Di Vecchia \cite{D'Adda:1978ur}.

The main result that we take away from this section is that in the
$\N=0$ theories where the fluctuation determinants contribute to the measure,
there is no $N_q$ dependence in the final result and so these terms 
play no r\^ole in the large-$N$ saddle-point approximation of the
measure undertaken in Sec.~4.

\subsection{The gauge invariant measure}

Since the measure is used to calculate the correlation
functions of gauge invariant operator insertions, it is convenient,
following \cite{MO-III}, to change variables to a set of gauge invariant
parameters and to explicitly integrate over the gauge 
degrees-of-freedom which parameterize a coset space. 
As explained
in \cite{MO-III}, this brings along a very significant advantage; namely
the non-linear bosonic ADHM constraints become trivial and can be explicitly
integrated out. 

A natural set of gauge-invariant collective coordinates, for
$N_q\geq2k_q$, is obtained by
constructing bosonic bilinear variables $W$ with diagonal blocks:
\begin{equation}
W^{(q)\dalpha}_{\phantom{(q)\dalpha}\dbeta}=\bar w^{(q)\dalpha}
 \,w^{(q)}_\dbeta\ .
\label{bosbi}\end{equation}
In terms of the $2K\times 2K$ matrix $W$, we then define
\begin{equation}
W^0={\rm tr}_2\,W,\quad W^c={\rm
tr}_2\,\tau^cW, \ \ c=1,2,3\,,
\label{aosbi}\end{equation}
where $W^0$ appeared previously in \eqref{Wdef}.

The gauge invariant measure then follows from the identity \cite{MO-III}
\begin{equation} 
\int_{\rm gauge\atop coset}dw^{(q)}\, d\bar
w^{(q)}\sim \big(\det_{2k_q}W^{(q)}\big)^{N_q-2k_q}
\, dW^{(q)0}\prod_{c=1,2,3}dW^{(q)c}\ ,
\label{E36.1}\end{equation}
valid for $N_q\geq2k_q$, applied to each diagonal block.
The constant of proportionality
was derived in Ref.~\cite{MO-III} but will not be needed for our present
purposes.
As already mentioned earlier, going to a gauge invariant measure brings 
a very significant
advantage: we can integrate out the ADHM $\delta$-functions explicitly. In
the bosonic sector, the ADHM constraints can be written succinctly in terms of
the gauge invariant variables as \cite{MO-III}
\begin{equation}
0= W^c - i [\,a'_n\,,\,a'_m\,]\,\etabar^c_{nm}\ .
\label{E27}\end{equation}
The remarkable feature of this re-writing of the constraints is that
it is {\it linear\/} in $W^c$ and consequently the $W^c$ integrals simply remove
the bosonic ADHM $\delta$-functions in \eqref{dmudef}.

There is a similar story in the fermionic sector. First of all we decompose
\begin{equation}
\mu^A=w_{\aD}\zeta^{\aD
A}+\nu^A,\qquad
\bar\mu^{A}=\bar\zeta
_\aD\bar w^{\aD}+\bar\nu^{A}\,.
\label{E44}\end{equation}
Here $\zeta^{\aD A}$ and $\nu^A$ 
have the same block diagonal form as
$\M^A$ as in \eqref{bfm}. So
$\zeta^{(q)\aD A}$ and 
$\bar\zeta^{(q)A}_\aD$ have dimension $k_q\times k_{q+q_A}$, while
$\nu^{(q)A}$ and $\bar\nu^{(q)A}$ have dimension $(N_q-2k_q)\times
k_{q+q_A}$ and $k_q\times (N_{q+q_A}-2k_{q+q_A})$, respectively. 
The variables $\nu^{A}$ lie in the orthogonal subspace to $w_\aD$, in the
sense that
\begin{equation}
\bar w^{\aD}\nu^{A}=0,\qquad \bar\nu^Aw^{}_{\aD}=0\ .
\label{E45}\end{equation}
In terms of the new variables the fermionic ADHM constraints
\eqref{adhmfc} are
\begin{equation}
\bar\zeta_{\bD}^AW_{\ \,\aD}^{\bD}+W_{\aD\bD}\zeta^{\bD A}+
\big[{\cal M}^{\prime\alpha A},a'_{\alpha\aD}\big]=0\ .
\label{E46}\end{equation}
The $\bar\zeta_\aD^A$ integrals then remove the fermionic ADHM
$\delta$-functions in \eqref{dmudef}.

Furthermore, in the fermionic sector we can
integrate out the $\nu^A$ and $\bar\nu^A$ variables; this being the
fermionic version of integrating over the gauge coset \cite{MO-III}.
In order to do this, it is
useful to split the fermion bilinear \eqref{newmatdef} that couples to $\chi_{AB}$ in
\eqref{E53} as
\begin{equation}
\Lambda^{AB}=\hat\Lambda^{AB}+\tilde\Lambda^{AB},
\label{E50}\end{equation}
where the first term has components just depending on $\{\nu^A,\bar\nu^A\}$:
\begin{equation}
\hat\Lambda^{AB}={1\over2\sqrt2}\left(\bar\nu^A\nu^B-
\bar\nu^B\nu^A\right),
\label{E51}\end{equation}
and the second term depends on the remaining variables
\begin{equation}
\tilde\Lambda^{AB}={1\over2\sqrt2}\big(\ 
\bar\zeta^A_\aD W^\aD_{\ \bD}\zeta^{\bD B}-\bar\zeta^B_\aD W^\aD_{\
\bD}\zeta^{\bD A}
+\{{\cal M}^{\prime\alpha A},{\cal M}^{\prime B}_\alpha\} \big)\ ,
\label{E52}\end{equation}
where $\bar\zeta^A_\aD$ is eliminated by the fermionic ADHM
constraints \eqref{E46}.
We can now explicitly integrate out the $\nu^A$'s and
$\bar\nu^A$'s. In general, unless $k_1=\cdots=k_p$, the result is
rather cumbersome to write down and for the large-$N$ calculation with $N\equiv
N_1=\cdots=N_p$, to
be undertaken in Sec.~4, we need only note
\begin{equation}
\int\prod_{A=1,2,3,4}d\nu^A\, d\bar\nu^A\,
\exp\big[\sqrt8\pi
ig^{-1}{\rm tr}_K
\,\chi_{AB}\bar\nu^A\nu^B\big]\sim\left({\rm det}_{
4K}\chi\right)^{N}\ .
\label{intoutnu}\end{equation}

\section{The Large-$N$ Instanton Measure}

In this section we investigate the ADHM gauge-invariant instanton measure
for the conformal field theories with $N\equiv N_1=\cdots=N_p$
in the large-$N$ limit by a steepest descent method following
Ref.~\cite{MO-III}. 
Our primary purpose is to determine the dependence of the large-$N$ measure
on the variables of the ADHM construction and so we
shall not keep track of the numerical pre-factor as we did in the
$\N=4$ theory.

\subsection{The saddle-point equations and their solution}

In order to take the large $N$ limit of the multi-instanton measure,
we have to gather together all the terms which involve the exponential
of a quantity times $N$. As explained in \cite{MO-III}, after rescaling
$\chi_{AB}\rightarrow\sqrt N\chi_{AB}$, there are three terms that
contribute to what can be viewed as an `effective action':
\begin{equation}S_{\rm
eff}=-2K(1+3\log2)-\log{\rm det}_{2K}W-\log{\rm det}_{4K}\chi +{\rm
tr}_K\,\chi_a\BL\chi_a\,.
\label{E57}\end{equation}
The second and third terms come from \eqref{E36.1} and \eqref{intoutnu} while the
final term comes from \eqref{E53}. Fortunately, as we have explained in
Sec.~3.3, the fluctuation determinants give an $N$ independent
contribution and so do not play any r\^ole in the saddle-point analysis.

We can now perform a steepest descent approximation of the measure
which involves finding the minima of the effective action with
respect to the variables $\chi_{AB}$, $W^0$ and $a'_n$. The resulting
saddle-point equations are\footnote{As usual we shall frequently
swap between the two representations $\chi_a$ and $\chi_{AB}$ for $SO(6)$ vectors.}
\begin{equation}
\begin{split}
\epsilon^{ABCD}\left(\BL\cdot\chi_{AB}\right) \chi_{CE}\ &=\
\tfrac12\delta^D_E \,1_{\sst [K]\times[K]}\ ,\\
\chi_a\chi_a\ &=\ \tfrac12(W^{-1})^0\ ,\\
[\chi_a,[\chi_a,a'_n]]\ &=\ i\bar\eta^c_{nm}[a'_m,(W^{-1})^c]\ .
\label{E60}
\end{split}
\end{equation}
where we have introduced the matrices
\begin{equation}
(W^{-1})^0=
{\rm tr}_2\,W^{-1},\qquad
(W^{-1})^c=
{\rm tr}_2\,\tau^cW^{-1}\ .
\label{E60.1}\end{equation}

We note that the expression for the effective action \eqref{E57}
and the saddle point equations \eqref{E60}
look identical to those derived in \cite{MO-III}
for the unprojected $\N=4$ theory. The solutions of these equations,
however, will not be the same as they have to be invariant under 
the ${\Bbb Z}_p$ projection. This means that 
$W^0$ and $a'_n$ have the block-diagonal
form \eqref{V30}, while $\chi_{AB}$ is generally
off-diagonal \eqref{V31}. With this in mind 
we shall use the analysis of the $\N=4$ theory as a guide and
draw heavily on the results derived in \cite{MO-III}. 
As in the $\N=4$ case we look for a solution with $W^c=0$,
$c=1,2,3$, which means that the instantons are embedded in mutually commuting
$\SU(2)$ subgroups of the gauge group. In this case 
equations \eqref{E60} are equivalent to
\begin{equation}
[a'_n,a'_m]=[a'_n,\chi_{AB}]=[\chi_{AB},\chi_{CD}]=0\ ,\quad
W^0=\tfrac12(\chi_a\chi_a)^{-1}\ .\label{echo}
\end{equation}
The final equation can be viewed as giving the value of $W^0$, the
instanton sizes, at the saddle-point and clearly $\chi_a\chi_a$ and
$W^0$ must be non-degenerate.

In the appendix we prove that in the type IIB case when $S^5/{\Bbb Z}_p$ is smooth 
there is no solution to
\eqref{echo} unless $k\equiv k_1=\cdots=k_p$. For the
analogous type 0B case, when $S^5/{\Bbb Z}_{p/2}$ is smooth, 
a solution only exists in the charge sector
with $k_+\equiv k_2=k_4=\cdots=k_p$ and $k_-\equiv k_1=k_3=\cdots=k_{p-1}$.
For the type IIB case, the 
general solution, up to the auxiliary symmetry, has the block-form
\begin{equation}\begin{split}
a^{\prime(q)}_n&={\rm
diag}\big(\,-\com^1_n\,,\,\ldots\,,\,-\com^k_n\,\big)\ ,
\\
\chi_{AB}^{(q)}&={\rm
diag}\big(\rho_1^{-1}\sfc_{AB}^{1}\,,\,\ldots\,,\,\rho_k^{-1}\sfc_{AB}^k\big)\ .\label{E61c}
\end{split}\end{equation}
Here $\sfc_{AB}^i$ are unit six-vectors:
\begin{equation}
\epsilon^{ABCD}\hat\Omega_{AB}^i\hat\Omega_{CD}^i=1\qquad\text{or}\qquad
\hat\Omega_a^i\hat\Omega_a^i=1\ ,
\label{sfive}\end{equation}
for each $i$. Notice that the discrete 
transformation $\sigma_{\{k,\ldots,k\}}\in\U(k)\times\cdots\times\U(k)$ 
fixes the form of \eqref{E61c} and implies some
discrete identifications of the $\sfc_{AB}^i$, explicitly $\sfc_{AB}^i\sim
e^{-2\pi i(q_A+q_B)/p}\sfc_{AB}^i$. So the coordinates $\sfc_{AB}^i$ are
valued on the quotient $S^5/{\Bbb Z}_p$.
The solution parameterizes the positions $\{\rho_i,X_n^i,\sfc_{AB}^i\}$
of $k$ D-instantons in $AdS_5\times S^5/{\Bbb Z}_p$. Notice that the
solution breaks the auxiliary symmetry to the maximal abelian subgroup
$\U(1)_{\rm diag}^k$ of the diagonal subgroup
$\U(k)_{\rm diag}\subset\U(k)\times\cdots\times\U(k)$.

Now we turn to the solution when $S^5/{\Bbb Z}_p$ has a
singularity. In this case solutions exist in all charge sectors
and there are multiple branches in the
solution space labelled by $\tilde k=0,\ldots,{\rm
min}(k_q)$:
\begin{equation}\begin{split}
a^{\prime(q)}_n&={\rm
diag}\big(\,-\com^1_n\,,\,\ldots\,,\,-\com^{\tilde
k}_n\,,\,-\com^{1,q}_n\,,\,\ldots\,,\,-X_n^{k_q-\tilde k,q}\,\big)\ ,
\\
\chi^{(q)}_{AB}&=\begin{cases}
{\rm
diag}\big(\,\rho_1^{-1}\sfc^1_a\,,\,\ldots\,,\,\rho_{\tilde k}^{-1}
\sfc^{\tilde k}_{AB}\,,\,\rho_{1,q}^{-1}\hat\Theta_{AB}^{1,q}\,,\,\ldots\,,\,
\rho_{k_q-\tilde k,q}^{-1}\hat\Theta_{AB}^{k_q-\tilde k,q}\,\big) &
q_A+q_B=0\ ,\\
{\rm
diag}\big(\,\rho_1^{-1}\sfc^1_a\,,\,\ldots\,,\,\rho_{\tilde k}^{-1}
\sfc^{\tilde k}_{AB}\,,\,0\,,\,\ldots\,,\,
0\,\big) & q_A+q_B\neq0\ .\end{cases}\label{E63c}
\end{split}\end{equation} 
In the above the $\hat\Theta_{AB}^{i,q}$ are unit vectors lying 
in $(S^5/{\Bbb Z}_p)_{\rm sing}$, in other words subject to
\eqref{sfive} but with $\Hat\Theta_{AB}^{i,q}=0$, for $q_A+q_B\neq0$.
In this case the solution represents $\tilde k$ D-instantons with
positions $\{\rho_i,X_n^i,\sfc_{AB}^i\}$ in $AdS_5\times S^5/{\Bbb
Z}_p$ along with $k_q-\tilde k$ fractional D-instantons of type $q$ with
positions $\{\rho_{i,q},X_n^{i,q},\hat\Theta_{AB}^{i,q}\}$ in
$AdS_5\times(S^5/{\Bbb Z}_p)_{\rm sing}$. The solution breaks the
auxiliary symmetry to $\U(1)_{\rm diag}^{\tilde
k}\times\U(1)^{K-p\tilde k}$. In the case $k\equiv
k_1=\cdots=k_p$ on the branch with $\tilde k=k$ we recover the solution
\eqref{E61c}. 

In the type 0B theories, the solutions are generalizations of
those above with the following difference. There are now two kinds of
D-instanton associated to the even and odd blocks. In particular,
when $S^5/{\Bbb Z}_{p/2}$ is smooth a solution only exist in the charge sector
$\{k_-,k_+,k_-,\ldots,k_+\}$. The fact that D-instantons now come in
two types matches precisely our expectation of the type 0B string theory.

\subsection{The large-$N$ instanton measure} 

In this section we construct the large-$N$ instanton measure. In
principle, we have to expand the effective action around the general solutions
written down in the last section to sufficient order 
to ensure that the fluctuation integrals converge. In
general because the Gaussian form has zeros whenever two D-instantons
coincide one has to go to quartic order in the fluctuations. 
Fortunately, as explained in \cite{MO-III}, 
we do not need to expand about the most general solution to the
saddle-point equations to quartic order since this is equivalent to expanding
to the same order around the most degenerate solution where all the 
D-instantons (in our present case both
fractional and non-fractional) are at the same point in
$AdS_5\times S^5/{\Bbb Z}_p$. 
The resulting quartic action has flat directions corresponding to the 
relative positions of the D-instantons. We can then recover the
original expansion that we wanted by re-expanding this action 
around some general point along a flat direction. 
It can then be established {\it ex post facto\/} that this is a consistent
procedure.

When $S^5/{\Bbb Z}_p$ has a singularity, the maximally
degenerate solution is \eqref{E63c} with all the
instantons at the same point. In particular, unless $k_1=\cdots=k_p$
there are fractional D-instantons and this means that only the components
$\chi_{AB}$ with $q_A+q_B=0$ are non-zero:
\begin{equation}
W^0=2\rho^21_{\sst[K]\times[K]}\ ,\quad
a'_n=-X_n1_{\sst[K]\times[K]}\
,\quad\chi_{AB}=\rho^{-1}\hat\Theta_{AB}1_{\sst[K]\times[K]}\ ,
\label{orbsp}
\end{equation}
where as before $\hat\Theta_{AB}$ is a unit vector in the directions 
fixed by ${\Bbb Z}_p$. The desired result for the large-$N$ measure is
immediate because on the saddle-point
solution \eqref{orbsp} all the variables are 
proportional to the identity matrix as in the $\N=4$ case and one can
essentially copy the analysis of Ref.~\cite{MO-III} verbatim.
The result is that the large-$N$ measure is simply, up to a normalization
constant, the partition function
of the ${\Bbb Z}_p$-projection of $\N=1$ 
supersymmetric ten-dimensional $\U(K)$ gauge theory dimensionally
reduced to zero dimensions. The ten-dimensional gauge field has
\begin{equation}
A_\mu=\big(\rho^{-1}a'_n,\rho\chi_{AB}\big)\ ,
\label{amudef}\end{equation}
and the ten-dimensional Majorana-Weyl fermion has components
\begin{equation}
\Psi=\sqrt{\pi\over2g}\,\big(\rho^{-1/2}{\cal
M}^{\prime A}_\alpha\,,\,
\rho^{1/2}\zeta^{\aD A}\big)\,.
\label{psidef}
\end{equation}
The ${\Bbb Z}_p$-projection acts as on the $\U(K)$-valued variables as 
\begin{equation}\begin{split}
\sigma_{\{k_q\}}a'_n\sigma_{\{k_q\}}^{-1}&=a'_n\ ,\quad
\sigma_{\{k_q\}}\chi_{AB}\sigma_{\{k_q\}}^{-1}=e^{2\pi i(q_A+q_B)/p}\chi_{AB}\ ,\\
\sigma_{\{k_q\}}\M^{\prime A}_\alpha\sigma_{\{k_q\}}^{-1}&=e^{-2\pi
iq_A/p}\M^{\prime A}_\alpha\ ,\quad
\sigma_{\{k_q\}}\zeta^{\aD A}\sigma_{\{k_q\}}^{-1}=e^{-2\pi
iq_A/p}\zeta^{\aD A}\ .
\end{split}\end{equation}
The partition function is then defined as the
integral over the projected variables of the ten-dimensional gauge
theory with an action where all the derivative terms are set to zero:
\begin{equation}
{\cal Z} =\  
\int_{\U(K);{\Bbb Z}_p} d^{10}A\, d^{16}\Psi\,e^{-NS(A_\mu,\Psi)}\ ,
\label{partf}\end{equation}
with
\begin{equation}
S(A_\mu,\Psi)\ =\  
-{1\over2}{\rm tr}_K\,\left[A_\mu,A_\nu\right]^2 
+N^{-1/2}{\rm
tr}_K\,\bar\Psi\Gamma_\mu\left[A_\mu,\Psi\right]\ .
\label{sukpart}
\end{equation}

The result for the large-$N$ measure when $S^5/{\Bbb Z}_p$ is smooth is
also the partition function of the corresponding ${\Bbb
Z}_p$-projected gauge theory, but only for the appropriate charge sectors,
i.e.~$\{k,\ldots,k\}$, for type IIB, and $\{k_-,k_+,k_-,\ldots,k_+\}$,
for type 0B. We consider the former case first.\footnote{The following 
analysis is also valid in
the same charge sector when $S^5/{\Bbb Z}_p$ has a singularity.}
The arguments leading to this conclusion are
more complicated because the maximally degenerate saddle-point
solution for $\chi_{AB}$ is not simply proportional to the identity. 
Below we sketch the proof.

First of all, the maximally degenerate solution is 
\begin{equation}
W^0=2\rho^21_{\sst[K]\times[K]}\ ,\quad
a'_n=-X_n1_{\sst[K]\times[K]}\ ,\quad\chi_{AB}=
\rho^{-1}\sfc_{AB}1_{\sst[k]\times[k]}\otimes\sum_{q=1}^pE_{q\,q-q_A-q_B}\ . 
\label{tbsp}
\end{equation}
Notice that $\chi_{AB}$, for $q_A+q_B\neq0$, is off-diagonal.
It is useful to introduce the notation $S_{AB}$ for the saddle-point solution
for $\chi_{AB}$. For each of the quantities
$v\in\{a_n',\chi_{AB}\}$ we define the following decomposition
\begin{equation}
v=v_0+\tilde v+\hat v\ ,
\label{decomp}\end{equation}
where $v_0$ is the saddle-point value of the variable, $\tilde v$ are
the components which do not commute with at least one $S_{AB}$.
The remaining variables $\hat v$ then commute with all the components $S_{AB}$.
Explicitly this means that
\begin{equation}
v_0=v'1_{\sst[k]\times[k]}\otimes\sum_{q=1}^p E_{q\,q+r}\ ,\quad
\tilde v=\sum_{q=1}^p\tilde v^{(q)}_{\sst[k]\times[k]}\otimes
E_{q\,q+r}\ ,\quad
\hat v=\hat v^{(0)}_{\sst[k]\times[k]}\otimes\sum_{q=1}^p E_{q\,q+r}\ .
\label{tdec}\end{equation}
with $\sum_{q=1}^p\tilde v^{(q)}_{\sst[k]\times[k]}=0$ and 
${\rm tr}_k\,\hat v^{(0)}_{\sst[k]\times[k]}=0$.

The complication in this case is that there are now
fluctuations that are lifted at Gaussian order.\footnote{These are in addition to the
fluctuations $\delta W^0$ that are lifted at this order: see
\cite{MO-III}.} The relevant terms in the expansion of the effective
action are
\begin{equation}
S^{(2)}=-{\rm tr}_K\,[S_a,\delta a'_n]^2-\rho^4{\rm
tr}_K\,[S_a,\delta\chi_b]^2
+\rho^4{\rm tr}_K[S_a,\delta\chi_a]^2\ .
\label{quadt}\end{equation}
So any fluctuation in $a'_n$ and $\chi_a$ 
that does not commute with $S_b$ will be lifted at
this order in the expansion. These are precisely the variables $\tilde
a'_n$ and $\tilde\chi_a$ in the decomposition \eqref{decomp}.

Actually, not quite all the
fluctuations $\tilde\chi_a$ are lifted by \eqref{quadt}
because there are exact flat directions of $S_{\rm eff}$ generated by
the action of the auxiliary $\U(k)\times\cdots\times\U(k)$ symmetry
group on the saddle-point solution. Infinitesimally, these are the variations
\begin{equation}
\tilde\chi^{\parallel}_a=[S_a,\epsilon]\,,
\end{equation}
where $\epsilon=\sum_{q=1}^p\epsilon^{(q)}\otimes E_{q\,q}$ are
infinitesimal parameters of the transformation. It is
convenient to `gauge-fix' this symmetry by integrating only
over fluctuations $\tilde\chi^\perp_a$ orthogonal to the variations
$\tilde\chi^\parallel_a$. These fluctuation can be specified by the conditions
\begin{equation}
{\rm tr}_K\,\big(\tilde\chi^\perp_a[S_a,\epsilon]\big)=0\,,\qquad\forall\epsilon\,,
\end{equation}
or equivalently $[\tilde\chi^\perp_a,S_a]=0$, hence the 
`gauge-fixed' quadratic action is therefore
\begin{equation}
S^{(2)}_{\rm gf}=-{\rm tr}_K\,[S_a,\tilde a'_n]^2-\rho^4{\rm
tr}_K\,[S_a,\tilde\chi^\perp_b]^2\,.
\label{quadgf}\end{equation}

Now we come to the crux of the argument. In general there could be cubic
interactions that couple one tilded variable with one two
hatted variables, i.e.~schematically of the form ${\rm tr}_K\,\tilde
v_1\hat v_2\hat v_3$, that contribute at the same order in $1/N$ as
the Gaussian terms in \eqref{quadgf} and quartic terms in the
hatted variables. However, because of the decompositions \eqref{tdec}
such couplings vanish. The quartic terms in the
hatted variables follow in an identical way to the $\N=4$
calculation \cite{MO-III}. Finally, at leading order in $1/N$ the bosonic variables
are controlled by the gauge-fixed action
\begin{equation}
S_{\rm gf}=-{\rm tr}_K\,[S_a,\tilde a'_n]^2-\rho^4{\rm
tr}_K\,[S_a,\tilde\chi^\perp_b]^2-{1\over2}{\rm
tr}_K\left(\rho^{-4}[\hat a'_n,\hat a'_m]^2+2[\hat\chi_a,\hat a'_n]^2
+\rho^4[\hat\chi_a,\hat\chi_b]^2\right)\, .
\label{halfw}\end{equation}
Up to a terms which are sub-leading in $1/N$, we can write this as a
gauge fixed version of the 
action of the ten-dimensional gauge field \eqref{amudef} as in
\eqref{sukpart}. The point is that \eqref{halfw} differs from
\eqref{sukpart} by (i) terms quadratic in tilded variables and
quadratic in hatted variables (ii) quartic in tilded variables. But
both these kinds of term are sub-leading in $1/N$. So in the resulting
partition function \eqref{partf} some of the terms in the action are
actually higher order in $1/N$. 
A similar result is follows in the $\{k_-,k_+,k_-,\ldots,k_+\}$ charge sector 
of the type 0B theories when $S^5/{\Bbb Z}_{p/2}$ is smooth.

In retrospect we can see the relation between the general solution of
the saddle-point equations and the flat directions of the gauge theory
action \eqref{sukpart}. The point is that the vanishing commutators
in \eqref{echo} are precisely the equations-of-motion of the gauge theory
action. In other words the space-of-solutions of the saddle-point equations is
identical to the `vacuum moduli space' of the matrix model. Therefore the
fact that the saddle-point solutions can be identified with the
moduli space of D-instantons is no accident. In the general case, 
the solutions describe the possible 
fractionations of D-instantons at the singularity $(S^5/{\Bbb
Z}_p)_{\rm sing}$. In the $\N=4$ case the
integrals over the relative positions of the D-instantons are
actually convergent and the charge $k$ D-instanton can be thought of
as a bound-state. It would be interesting to investigate the
convergence of the integrals over the relative positions of the
(fractional) D-instantons in these more general theories. In
particular, the convergence issue is crucial for finding the 
$N$-dependence of the large-$N$ instanton measure.

\subsection{Contribution to 16 fermion correlators} 

In the $\N=4$ theory, the simplest correlation functions which
received instanton contributions involved insertions which only saturate
the Grassmann integrals over the collective coordinates corresponding
to the 16 supersymmetric and superconformal zero-modes. In this
section, we show that the effective large-$N$ measure in the charge
sector $\{k,\ldots,k\}$ in the analogous fermionic sector is
identical to the $\N=4$ case. By analogous fermionic sector we mean
the sector where the integrals over the 16 variables $\xi^A_\alpha$
and $\bar\eta^{\aD A}$ defined by 
\begin{equation}
\M^{\prime
A}_\alpha=\xi^A_\alpha1_{\sst[k]\times[k]}\otimes\sum_{q=1}^pE_{q\,q-q_A}\
,
\qquad\zeta^{\aD A}=\bar\eta^{\aD
A}1_{\sst[k]\times[k]}\otimes\sum_{q=1}^pE_{q\,q-q_A}\ ,
\label{susyzm}\end{equation}
are saturated by insertions rather than from terms in the action of
the partition function \eqref{partf}. 
These variables obviously include the collective coordinates $\xi^A_\alpha$
and $\bar\eta^{\aD A}$, for $q_A=0$,
corresponding to the supersymmetric and superconformal zero-modes
modes.\footnote{Actually the definition of the supersymmetric and
superconformal collective coordinates differs from \eqref{susyzm} by a
linear shift which we may transform away: see \cite{MO-III}.}

So the effective measure in this fermionic sector is defined by
separating out the integrals over $\xi^A_\alpha$ and $\bar\eta^{\aD
A}$ from \eqref{partf} and leaving them unsaturated. The appropriate
thing to do now, and which will be justified {\it ex post facto\/}, 
is to expand the action around the branch in the
solution space in \eqref{tbsp}. The gauge fixing described in the last
section leading to the action \eqref{halfw} is valid in this context. 
We can integrate the
fluctuations lifted at Gaussian order $\tilde a'_n$ and
$\tilde\chi_a^\perp$:
\begin{equation}
\int d\tilde a'\,d\tilde\chi^\perp\,\exp\Big(-N{\rm tr}_K\,[S_a,\tilde a'_n]^2-N\rho^4{\rm
tr}_K\,[S_a,\tilde\chi^\perp_b]^2\Big)=\rho^{-10k^2(p-1)}
N^{-9k^2(p-1)/2}\big(\det'_pM\big)^{-9k^2/2}\ ,
\label{gint}\end{equation}
where $M$ is the $p\times p$ matrix\footnote{It is useful
to introduce a basis of vectors to match the block-form
of the matrices. To this end we define the $p$-dimensional column
vector $e_q$ with a 1 in the $q^{\rm th}$ position and 0
elsewhere. These vectors have the property 
$E_{q\,r}e_s=e_q\delta_{rs}$.}
\begin{equation}
M=2\sum_{q=1}^p\epsilon^{ABCD}\hat\Omega_{AB}\hat\Omega_{CD}(E_{q\,q-q_A-q_B}-
E_{q\,q})\,.
\label{defm}\end{equation}
The prime on the determinant in \eqref{gint} implies that we should 
take a product only over the non-null eigenvalues of $M$ since the latter
has a null eigenvector of the form
$\sum_{q=1}^pe_q$. In addition, we must take into account the
gauge-fixing Jacobian:
\begin{equation}
J_{\rm GF}=({\rm Vol}\,U(k))^{p-1}
\rho^{-(p-1)k^2}\big(\det'_p\,M\big)^{k^2/2}\ .
\label{jacob}\end{equation}
To arrive at our answer, we must also integrate-out the fermionic
partners to the Gaussian variables, namely, $\tilde\M^{\prime
A}_\alpha$ and $\tilde\zeta^{\aD A}$. These are coupled, to the
saddle-point solution via the terms 
\begin{equation}
NS_{\rm f}=\Big({2\pi^2N\over g^2}\Big)^{1/2}\,
{\rm tr}_K\,\big(\rho[S_{AB},\tilde\zeta^{\aD
A}]\tilde\zeta^B_\aD+\rho^{-1}[S_{AB},\tilde{\cal M}^{\prime
\alpha A}]\tilde{\cal M}^{\prime B}_\alpha\big)\ .
\label{messa}\end{equation}
The fermionic integrals are then 
\begin{equation}
\int \prod_{A=1,2,3,4}d\tilde\zeta^A\,d\tilde{\cal M}^{\prime
A}\,e^{NS_{\rm f}}=\left({2\pi^2N\over
g^2}\right)^{4(p-1)k^2}(\det'_{4p}G)^{2k^2}\ ,
\label{fermint}\end{equation}
where $G$ is the $4p\times4p$ matrix with block-form elements 
\begin{equation}
G_{AB}=\sfc_{AB}\sum_{q=1}^p(E_{q\,q-q_A-q_B}-E_{q\,q})
\label{defg}\end{equation}
and, as before, the prime, indicates the removal of the zero eigenvalue.
Since as $p\times p$ matrices $[G_{AB},G_{CD}]=0$, it is straightforward to show that
\begin{equation}
\det'_{4p}G=\big(\det'_p(\tfrac18\epsilon^{ABCD}G_{AB}G_{CD})\big)^2=
\big(\det'_p\tfrac18M\big)^2\ ,
\end{equation}
where $M$ is the matrix defined in
\eqref{defm}. Notice that the determinant factors of $M$ 
cancel between \eqref{gint}, \eqref{jacob} and \eqref{fermint}.

What remains are the integrals over the quartic fluctuations $\hat
a'_n$ and $\hat\chi_a$ along with their fermionic partners which have
the form of the partition function of ten-dimensional $\N=1$
supersymmetric $\SU(k)$ gauge theory dimensionally reduced to zero
dimensions. Being careful with the factors of $N$ and $g$, our final
result for the large-$N$ measure in this sector is
\begin{equation}
\int d\mu_{\rm phys}^{\{k,\ldots,k\}}e^{-S_{\rm inst}}
\ \underset{N\rightarrow\infty}=\  
g^8\sqrt Ne^{2\pi
ikp\tau}\int{d^4X\,d\rho\over\rho^5}\,d^5\sfc\,\prod_{A=1,\ldots,4}\,
d^2\xi^A\,d^2\bar\eta^A\cdot{\cal Z}_{\SU(k)}+\cdots\ .
\label{fmes}\end{equation}
where the ellipsis reminds us that the expression is only the part of
the measure where the 16 variables $\xi_\alpha^A$ and $\bar\eta^{\aD
A}$ are left un-integrated. 

Notice that \eqref{fmes} is identical to the large-$N$ measure in the
charge $k$ sector of the $\N=4$ theory with $\tau\rightarrow
p\tau$. Hence charge $k$ instanton contributions to the 16 fermion correlators
evaluated in the $\N=4$ theory will be identical to those from the
charge $\{k,\ldots,k\}$ sector in the orbifold theories.

\section{Relation to $\N=2$ $\SU(N)$ Gauge Theory with $N_F=2N$}

In the above we have considered the instanton contributions in the
large-$N$ limit to a class of $\N=2$ theories with
product group structure $\SU(N)\times\cdots\times\SU(N)$ (after
decoupling the abelian factors). In Ref.~\cite{N=2} we considered the
contributions of instantons in the large-$N$ limit to the $\N=2$
conformal theory with gauge group $\SU(N)$ and with $N_F=2N$ fundamental
hypermultiplets. Although apparently unconnected, the instanton
contributions in these theories are related in the following way. 
Consider the charge sector $\{k,0,\ldots,0\}$ in the $\N=2$
supersymmetric theories with gauge group 
$\SU(N)\times\cdots\times\SU(N)$. In this instanton background
only the $\SU(N)$ gauge field $v_n^{(1)}$ is non-zero, and
consequently only the adjoint fermion fields $\lambda^{(1)A}$, with $A=1,2$,
and adjoint scalar fields $A^{(1)AB}$, with $AB=12,34$,
along with the fermion fields $\lambda^{(q)A}$, with $q=1,p$ and $A=3,4$,
and the scalar
fields $A^{(q)AB}$, with $q=1,p$ and $AB=13,14,23,24$, which
transform in either the $\BN$ or $\bar\BN$ representations of the
$\SU(N)$ factor, are non-zero.
These fields amass into the an $\SU(N)$  
vector representation of $\N=2$ supersymmetry and $2N$ fundamental
hypermultiplets; precisely the theory considered in Ref.~\cite{N=2}.
So the instantons in the  $\{k,0,\ldots,0\}$ charge sector of the
first theory should describe the charge $k$ instantons of the
second theory.

Now we are in a position to interpret the results of Ref.~\cite{N=2}
for the large-$N$ instanton measure in the context of the results in
this paper. In the charge sector $\{k,0,\ldots,0\}$ the
large-$N$ instanton measure involves the partition function for the
${\Bbb Z}_p$-projected
$\N=1$ supersymmetric $\U(k)$ gauge theory in ten-dimensions
dimensionally reduced to zero dimensions, with ${\Bbb Z}_p$
embedded in the gauge group as
$\sigma_{\{k,0,\ldots,0\}}=1_{\sst[N]\times[N]}$ and in $\SU(4)_R$ as
in \eqref{gsu}. Only the $\U(k)$-adjoint matrices 
$\chi_{AB}^{(1)}$, $AB=12,34$, ${\cal
M}^{\prime(1)A}_\alpha$ and $\zeta^{(1)\aD A}$, with $A=1,2$, survive this
projection. The resulting partition function can equally well be described as
that of the $\N=(1,0)$ supersymmetric $\U(k)$
theory in six dimensions dimensionally reduced to zero
dimensions. This is precisely the result of Ref.~\cite{N=2}. In
particular since from the orbifold perspective all the $k$
instantons are fractional we have a simple explanation of the
observation that the saddle-point solution of  Ref.~\cite{N=2}
described a point-like object in $AdS_5\times S^1$: the $S^1$ is
precisely the orbifold singularity of $S^5/{\Bbb Z}_p$ on which the
fractional D-instantons are confined to move.

\section{Discussion}

In the $\N=4$ theories much more could be achieved \cite{MO-III} because
the measure could be used to calculate
instanton contributions to particular correlation functions which
could then be compared with D-instanton induced terms in the type IIB
supergravity effective action \cite{Green:rev}. In this comparison the overall
numerical pre-factor in the large-$N$ ADHM measure was crucial to get
the precise agreement found in \cite{MO-III}. For the orbifold
theories we have not kept track of the numerical pre-factor since it
is not clear on the dual side what to compare with. In fact a dual
supergravity approximation of the string theory only exists if
$S^5/{\Bbb Z}_p$ is smooth, unless the singularity is blown up in some
way. It would clearly be interesting to attempt the more detailed
check of the duality carried through in \cite{MO-III}.

\acknowledgments

We thank Lance Dixon, Nick Dorey, Michael Green, Igor Klebanov
and Michael Mattis for discussions and comments. 
We also acknowledge the TMR network grant FMRX-CT96-0012.

\startappendix

\rsen
\Appendix{}
In this appendix we prove that solutions to the saddle-point
equations \eqref{echo} for the type IIB dual cases when
$S^5/{\Bbb Z}_p$ is smooth only exist if all the $k_q$
all equal.

Our proof proceeds along the following lines.
Since all the matrices $\chi_{AB}$ commute they have a basis of
simultaneous eigenvectors. In order to show that there are no
solutions to \eqref{echo} unless all the instanton
numbers in each group factor are the same 
we will show that in the
contrary situation the matrices $\chi_{AB}$ have at least one common
null eigenvector and consequently \eqref{echo} cannot be satisfied
because $W^0$ is non-degenerate. 

Consider, for the moment, one of the $\chi_{AB}$ matrices, which we
denote generically as $A$, with the block form
\begin{equation}
A=\sum_{q=1}^pA^{(q)}\otimes E_{q\,q-r}\, .
\label{}\end{equation}
Since ${\Bbb Z}_p$ acts freely on $S^5$ we have $r\neq0$
mod $p$. In general, if  $r$ and $p$ have some 
common integer divisor(s) (the largest of which we denote $c$)
${\Bbb Z}_p$ does not act faithfully on $A$, rather only the subgroup
${\Bbb Z}_{p/c}$ is realized.

The first result we need, is that any non-null
eigenvector of $A$ has the form
\begin{equation}
v=\sum_{q=1}^{p/c}v^{(rq+s)}\otimes e_{rq+s}\, ,
\label{forme}\end{equation}
where all the components $v^{(rq+s)}$, $q=1,\ldots,p/c$, are all
non-vanishing. To see this consider the eigenvalue equation
in the block-form basis:
\begin{equation}
A^{(r+s)}v^{(s)}=\lambda v^{(r+s)}\,.
\label{}\end{equation}
Iterating this equation $p/c$ times we cycle back to yield an
eigenvector equation for
$v^{(s)}$:
\begin{equation}
A^{(rp/c+s)}\cdots A^{(r+s)}v^{(s)}=\lambda^{p/c}v^{(s)}\,.
\label{}\end{equation}
Denoting the eigenvalue as $\mu$, we have $\lambda=\mu^{c/p}$. This
means that each non-null eigenvector of $A^{(rp/c+s)}\cdots
A^{(r+s)}$ gives rise to $p/c$ non-null eigenvectors of $A$ itself,
the multiplicity arising from the 
$p/c$ choices of the roots of $\lambda=\mu^{c/p}$. The non-null
eigenvectors of $A$ are consequently of the form
\begin{equation}
v=\sum_{q=1}^{p/c}\mu^{-cq/p}A^{(rq+s)}\cdots 
A^{(r+s)}v^{(s)}\otimes e_{rq+s}\, .
\label{}\end{equation}
As stated,
the non-null eigenvectors have the form \eqref{forme} where all
the components $v^{(rq+s)}$, $q=1,\ldots,p/c$, are all non-vanishing.

The fact that the non-null eigenvectors
have the form \eqref{forme} implies that all the remaining eigenvectors
must be null. The null eigenvectors can therefore be written as 
\begin{equation}
v^{(q)}\otimes e_q\,,
\label{formn}\end{equation}
with no sum on $q$. A lower bound 
on the number of null eigenvectors of the form \eqref{formn} is consequently
\begin{equation}
d(r,q)=k_q-k_{\sst\rm min}(r,q)\,,
\end{equation}
where $k_{\sst\rm min}(r,q)$
is the smallest number in the set $\{k_{q+nr},\ n=1,\ldots,p/c\}$.

We now denote the first matrix as $A_1$, with associated quantities 
$r_1$ and $c_1$, and introduce a second matrix
$A_2$ of the same form:
\begin{equation}
A_2=\sum_{q=1}^pA_2^{(q)}\otimes E_{q\,q-r_2}\, 
\end{equation}
that commutes with $A_1$. We now proceed to establish a lower bound
on the number of simultaneous null eigenvectors of $A_1$ and $A_2$. The
point is that certain linear combinations of null eigenvectors of $A_1$
of the form can be non-null eigenvectors of $A_2$, and
vice-versa. We must enumerate the maximum possible such combinations
that can arise to establish a lower bound on the simultaneous null
eigenvectors of $A_1$ and $A_2$. Consider the null eigenvectors of
$A_1$ of the form \eqref{formn}. A combination of such null
eigenvectors of the form
\begin{equation}
v^{(q)}\otimes e_{q}+
v^{(q+r_2)}\otimes e_{q+r_2}+\cdots+v^{(q+(p/c_2-1)r_2)}\otimes e_{q+(p/c_2-1)r_2}\ ,
\end{equation}
can be a non-null eigenvector of $A_2$. The number of such
combinations that have this property is clearly bounded above by
\begin{equation}
{\rm min}\big(d(r_1,q),d(r_1,q+r_2),
\ldots,d(r_1,q+(p/c_2-1)r_2)\big)\ .
\end{equation}
Hence a lower bound on the number of simultaneous null eigenvectors  
of the form $v^{(q)}\otimes e_q$ is given by
\begin{equation}\begin{split}
d(r_1,q)-&{\rm min}\big(d(r_1,q),
\ldots,d(r_1,q+(p/c_2-1)r_2)\big)\\
&={\rm max}\big(0, d(r_1,q)-
d(r_1,q+r_2),\ldots,d(r_1,q)-d(r_1,q+(p/c_2-1)r_2)\big)
\end{split}\end{equation}
However, we can also run the argument the other way and consider the
null eigenvectors of $A_2$ that can be non-null eigenvectors of
$A_1$. Taken together this gives a lower bound $X_q$ on the number of
simultaneous null eigenvectors of $A_1$ and $A_2$ of the form $v^{(q)}\otimes e_q$
\begin{equation}
X_q={\rm max}\big(d(r_1,q)-d(r_1,q+n_2r_2),
d(r_2,q)-d(r_2,q+n_1r_1)\big)\,.
\end{equation}
The expression above is written in short-hand
where the integers run over the values $n_{1,2}=0,\ldots,p/c_{1,2}-1$.

It is now easy to introduce a third matrix $A_3$ and establish a lower
bound on the simultaneous null eigenvectors of all three matrices. The
idea is to consider the maximum number of simultaneous null
eigenvectors of $A_1$ and $A_2$ that can be non-null eigenvectors of
$A_3$. For null eigenvectors of the form $v^{(q)}\otimes e_q$, this
number is bounded above by 
\begin{equation}
{\rm min}\big(X_q,X_{q+r_3},
\ldots,X_{q+(p/c_3-1)r_3}\big)\ .
\end{equation}
Hence a lower bound on the number of simultaneous null eigenvectors of
$A_1$, $A_2$ and $A_3$ of the form $v^{(q)}\otimes e_q$ is given by
\begin{equation}\begin{split}
{\rm max}\big(0,X_q-X_{q+r_3},\ldots,X_q-X_{q+(p/c_3-1)r_3}\big)\,.
\end{split}\end{equation}
This establishes a lower bound on the number of simultaneous
null eigenvectors of all three matrices. The most stringent bound is
then obtained by considering all three permutations of this line of
reasoning. Putting all this together, one finds a lower bound
\begin{equation}
{\rm max}\big(d(r_a,q)-d(r_a,q+n_br_b)-d(r_a,q+n_cr_c)+d(r_a,q+n_br_b+n_cr_c)\big)
\label{lbo}\end{equation}
on the number of simultaneous null eigenvectors,
where $a,b,c$ run over the 6 permutations of $1,2,3$, and the integers
$n_a=0,\ldots,p/c_a-1$.

Using the equations \eqref{echo} for $\chi_{AB}$ and the
reality condition \eqref{E54}, the last equation in \eqref{echo} can be written as
\begin{equation}
\chi_{12}^\dagger\chi_{12}+\chi_{13}^\dagger\chi_{13}+\chi_{14}^\dagger\chi_{14}
=\tfrac14(W^0)^{-1}\,.
\label{rwecho}\end{equation}
If $\chi_{12}$, $\chi_{13}$ and $\chi_{14}$ have a simultaneous null
eigenvector then there can be 
no solution to this equation since $W^0$ is non-degenerate 
But \eqref{lbo},
with $r_1=q_1+q_2$, $r_2=q_1+q_3$ and $r_3=q_1+q_4$, gives a lower
bound on the number of such null eigenvectors; hence for a solution \eqref{rwecho} to
exist, it is necessary that for each $q$ and permutation of $a,b,c$ 
\begin{equation}
d(r_a,q)-d(r_a,q+n_br_b)-d(r_a,q+n_cr_c)+d(r_a,q+n_br_b+n_cr_c)\leq0\,.
\end{equation}
The only way this set of inequalities can be
satisfied is if $d(r_a,q+n_cr_c)-d(r_a,q+n_br_b+n_cr_c)$ is
independent of $n_c$ which further requires that all the $k_q$ have to be equal.
{\it QED\/}.

\end{document}